\documentclass{aa}  
\pdfoutput=1 
\usepackage{hyperref}
\usepackage{graphicx}
\usepackage{txfonts}
\usepackage{amsmath}
\usepackage{float}
\usepackage{physics}
\usepackage{siunitx}
\usepackage[dvipsnames]{xcolor}

\newcommand{\vc}{\ensuremath{v_{\mathrm{c}}(R)}}
\newcommand{\Gaia}{\textit{Gaia}}

\newcommand{\kpc}{\ensuremath{\,\mathrm{kpc}}}
\newcommand{\kms}{\ensuremath{\,\mathrm{kms}^{-1}}}
\newcommand{\Auriga}{\texttt{Auriga}}
\defcitealias{Ou2023}{O23}

\begin{document} 

%---------------------------------------------
%                  Content
%---------------------------------------------

%---------------------------------------------
%                   Title
%---------------------------------------------

\title{On the Galactic rotation curve inferred from the Jeans equations}
\subtitle{Assessing its robustness using \textit{Gaia} DR3 and cosmological simulations}

%---------------------------------------------
%                   Authors
%---------------------------------------------

 \author{{Orlin Koop\inst{1}\fnmsep\thanks{e-mail: koop@astro.rug.nl}}
          \and 
          {Teresa Antoja}
          \inst{2}
          \and 
          {Amina Helmi}
          \inst{1}
          \and 
          {Thomas M. Callingham}
          \inst{1}
          \and
          {Chervin F. P. Laporte}
          \inst{2}
          }

   \institute{Kapteyn Astronomical Institute, University of Groningen, Landleven 12, NL-9747 AD Groningen, the Netherlands
   \and
   Institut de Ci\`encies del Cosmos (ICCUB), Universitat de Barcelona (IEEC-UB), Mart\'i Franqu\`es 1, 08028 Barcelona, Spain
   }

   \date{Received xxxx; accepted yyyy}

%---------------------------------------------
%                   Abstract
%---------------------------------------------
\abstract
  % context heading (optional)
   {Several works have recently applied Jeans modelling to \Gaia{}-based datasets to infer the circular velocity curve for the Milky Way. Such works have consistently found evidence for a continuous decline in the rotation curve beyond $\sim15~\si{kpc}$ possibly indicative of a light dark matter halo.}
  % aims heading (mandatory)
  {Using a large sample from \textit{Gaia} DR3 data we aim to derive the rotation curve of the Milky Way using the Jeans equations and, in particular, 
   to quantify the role of systematic effects, both in the data and inherent to the Jeans methodology under the assumptions of axisymmetry and time-independence.
   }
  % methods heading (mandatory)
   {We have used data from the \textit{Gaia} DR3 RVS sample, supplemented with distances inferred through Bayesian frameworks to determine the radial variation of the second moments of the velocity distribution for stars close to the Galactic plane. We have used these profiles to determine the rotation curve using the Jeans equations under the assumption of axisymmetry and explored how they vary with azimuth and above and below the Galactic disk plane. 
   We have applied the same methodology to an N-body simulation of a Milky Way-like galaxy impacted by a satellite akin the Sagittarius dwarf and to the  \texttt{Auriga} suite of cosmological simulations.}
  % results heading (mandatory)
   {The circular velocity curve we infer for the Milky Way is consistent with previous work out to $\sim 15$~kpc, where our statistics are robust. Due to the larger number of stars in our sample, we are able to reveal 
   evidence of disequilibrium and deviations from axisymmetry closer in. For example, we find that the second moment of $V_R$ flattens out at $R \gtrsim 12.5$~kpc,
 and that the second moment of $V_\phi$ is different above and below the plane for $R \gtrsim 11$~kpc.
Our exploration of the simulations indicates that these features are typical of galaxies that have been perturbed by external satellites. Also from the simulations we estimate that the difference between the true circular velocity curve and that inferred from Jeans equations can be as high as $15\%$, but that it likely is of order $10\%$ for the Milky Way. This is of larger amplitude than the systematics associated to the observational uncertainties or those from most modelling assumptions when using the Jeans equations. However, if the density of the tracer population were truncated at large radii instead of being exponential as often assumed, this could lead to the erroneous conclusion of a steeply declining rotation curve.}
    % Conclusions
   { We find that steady-state axisymmetric Jeans modelling becomes less robust at large radii,
    indicating that particular caution is needed when interpreting the rotation curve inferred in those regions.
    A more careful and sophisticated approach may be necessary for precision measurements of the dark matter content of our Galaxy.   
    }

%---------------------------------------------
%                   Keywords
%---------------------------------------------

\maketitle

\section{Introduction} \label{sec:intro}

A galaxy's circular velocity curve, denoted \vc{}, provides a measure of its gravitational potential,
which is related to the mass distributions of both the visible baryonic structure and hidden dark content. Studies of extragalactic rotation curves based on HI gas motions, have, throughout the years, provided strong evidence for the `missing mass' that had puzzled Zwicky half a century earlier 
\citep[][and more recently see also \citealt{deBlok2008,Lelli2016}]{Zwicky1933,Rubin1980,Bosma1981}. Therefore, rotation curves have arguably helped place dark matter (DM) at the centre of modern astrophysics. 

The currently preferred cosmological paradigm, $\Lambda$CDM, predicts that visible galaxies are embedded in a DM halo, distributed following an NFW density profile \citep{NFWprofile}.
This functional form however, does not always match the diversity of rotation curves of (dwarf) galaxies, sometimes even showing an unexpected decline at large radii \citep{NFWprofile,deBlok2008,Oman2015,Genzel2017}. Additionally, a large diversity in rotation curve shapes can be expected in external galaxies \citep{Sands2024}.
High-redshift measurements of rotation curves 
by \cite{Kretschmer2021} and \cite{Roman-Oliveira2023} also show that observations do not align with predictions from $\Lambda$CDM-driven simulations. Since feedback processes can strongly influence the DM distribution of inside galaxies \citep{Read2016,ManceraPina2022}, these findings need to be interpreted with some care. It is clear that 
circular velocity curves serve to explore the influence of baryonic processes on the mass distribution in a galaxy. They also can be used to test    alternative theories of dark matter or gravity \citep{Kamada2017,Banares-Hernandez2023,Petersen2020}. 
%%% MW
In our own galaxy, the Milky Way (MW),
understanding the Galactic mass distribution is critical to understanding the dynamics that govern our surroundings.
Reliably modelling our \vc{} requires accurate kinematics for a sizable sample of stars.
From our position within the Galaxy, we have extremely high-quality local data.
However, our view of the wider Galaxy is severely hindered at greater distances in both quantity and quality.
Early investigations of the MW's \vc{} used measurements of HI in the inner regions of our Galaxy
and of standard candles like Cepheids and Red Clump giants in the outer disk 
\citep{Pont1997,Bovy2012}. In combination with tracers outside of the disk, such as 
RR Lyrae, Blue Horizontal Branch stars and globular clusters \citep{Xue2009,Ablimit2017}, these measurements have been used to infer the enclosed mass profile, or in other words, a "unified" circular velocity curve  \citep[][see also the more recent \citealt{Sofue2020}]{Sofue2009}.

%% Gaia
The release of \Gaia{} DR2 significantly improved the quality and extent of the data,
containing astrometry and radial velocity measurements for $\sim7.2$ million stars in their radial velocity spectrograph (RVS) catalogue.
This dataset was used as a base by several works to estimate an accurate Galactic rotation curve,
attempting to reach larger radii than previously possible \citep{Mroz2019,Eilers2019}.
To improve distance estimates,
the \Gaia{} dataset was complimented with other surveys;
e.g. \citet{Eilers2019} used  \textit{APOGEE} spectrometry \citep{Majewski2017},
whilst  \citet{Mroz2019} used a catalogue of Cepheids \citep{Skowron2019_Cephids}.
In the \vc{} inferred by \citet{Eilers2019}, the errors increase noticeably past $\sim20\kpc$,
where some works truncate it for fitting purposes \citep[e.g.][]{Cautun2020}.
The resulting fits approximately matched what is expected for an NFW profile
and agreed with a slowly improving consensus on the total mass of the Milky Way of $\left[0.5-1.5\right]\times10^{12}$
\citep{Callingham2019,Wang20_MWMass}.

%Gaia DR3 and BEyon
In the latest \Gaia{} data release, DR3 \citep{GaiaeDR3,GaiaCollaboration2023DR3}, the RVS catalogue contains 34 million stars,
and parallax uncertainties have been reduced by a factor of 1.25.
With this data, a new selection of works have applied previously developed techniques to infer distances for stars at large radii (even out to 30 kpc) and to fit the \vc{} \citep{Ou2023,Zhou2023, Wang2023, Jiao2023}. The works of \citet{Ou2023} and \citet{Zhou2023} follow \citet{Eilers2019},
using a linear model that combines spectroscopic data from \textit{APOGEE} (and \textit{LAMOST}) with photometry from \textit{Gaia} to estimate reliable spectrophotometric distances for
Red Giant Branch stars \citep[RGBs,][]{Hogg2019}.
Alternatively, \citet{Wang2023} used a statistical tool known as the Lucy Inversion Method \citep{Lucy1974_LIM}
to estimate distances and velocity dispersions for populations of stars binned in  6D phase-space cells.
All these works find reasonably consistent results with a declining \vc{}, and favor a lighter mass of the MW when extrapolated outwards. For example
\citet{Ou2023} suggest that  $M_{200}$ may be as low as $6.9\times10^{11}M_{\odot}$ (for a generalised NFW halo, which however provides a suboptimal fit), comparable to the estimated $M_{200} \sim 8 \times 10^{11} M_\sun$ by \citet{Zhou2023}, while 
\citet{Jiao2023} claim that the decline seen would be consistent with the expectation of a Keplerian fall-off. 

%Jeans
These studies all derive $v_{\rm c}(R)$ for the Milky Way using the axisymmetric Jeans equations, which link the moments of the velocity distributions of a tracer population and the gravitational potential of an axisymmetric and time-independent stellar system.
However, it is known that the Milky Way is not entirely axisymmetric or in dynamical equilibrium.
Signs of this include spiral arm structure \citep[e.g. in][]{GaiaCollaboration2023Drimmel}, the discovery of the north-south asymmetry in the kinematics of disk stars \citep[e.g. in][]{Widrow2012,Williams2013,GaiaCollaboration2021anti},
the phase-space spiral \citep{Antoja2018}, and the interaction of the Milky Way with the Magellanic Clouds and Sagittarius \citep[see e.g.][]{VasilievTango2021}. 

It is hard to predict  analytically the impact of these deviations upon the derived \vc{}. 
Fortunately, it is possible to study these effects through simulations of MW-like analogues.
Already, cosmological simulations like FIRE \citep{ElBadry2017}, \texttt{SURFS} \citep{Kafle2018}, APOSTLE \citep{Wang2018} and VELA \citep{Kretschmer2021}
have shown that applying the spherical Jeans equations to DM halos or tracer stellar particles in those halos can cause
a bias in the estimation of the virial mass of the system of up to $20\%$.
In a study of the axisymmetric Jeans equations applied N-body simulations,
\cite{Chrobakova2020} concluded that if the amplitude of the mean radial motions (expected to be zero in equilibrium and axisymmetry) are sufficiently small compared to the rotational motion, the results are likely reliable. The work of \cite{Haines2019} explored the impact of recent (massive) satellite passage on the vertical Jeans equations in the N-body simulation L2 of a MW-like stellar disk by \citet{Laporte2018}, concluding that the heavy perturbations on the disk affect traditional Jeans modelling, particularly in the underdense areas at large radii.

%This Work
In this work, we calculate the MW's circular velocity curve with a large \textit{Gaia} DR3 RVS sample of RGB stars,
using calibrated distances from \citet{Bailer-Jones2021} and StarHorse \citep{Anders2022},
and estimates of stellar parameters from \citet{Andrae2023}.
These datasets are an order of magnitude larger than previous \textit{APOGEE} based work, although the range of distances they probe is smaller. As in previous works, we explore the systematic uncertainties associated with the different terms in the axisymmetric Jeans equations.
We then use numerical simulations to study time-dependent effects and deviations from axisymmetry, especially from interactions with satellites,
to further understand the intricacies and limitations of Jeans modelling on the Milky Way.
We consider an N-body MW-like galaxy being impacted by a Sagittarius-like dwarf galaxy \citep[L2 from][]{Laporte2018}.
Additionally, we use the {\tt Auriga} suite of cosmological simulations \citep{Grand2017} to study Milky Way-like galaxies in a cosmological environment,
including their associated {\tt Aurigaia} \Gaia{} DR3 mock catalogues to understand the effects of observational errors \citep{Grand2018aurigaia}.

%Paper
This paper is structured as follows.
Section \ref{sec:data} describes our data samples.
Section \ref{sec:methods} presents the Jeans equations and analyses the individual terms for our dataset.
In Section \ref{sec:rotation curve}, we calculate the MW's rotation curve and discuss the important systematic uncertainties.
In Section \ref{sec:simulations:Sagittarius}, we apply our methodology to the L2 simulation of a MW-like disk interacting with a Sagittarius-like dwarf galaxy.
Section \ref{sec:simulations:auriga} describes applying Jeans modelling to the {\tt Auriga} suite, and their corresponding {\tt Aurigaia} mock catalogues.
In Section \ref{sec:discussion}, we discuss our results and the implications for Jeans modelling within the MW.
Finally,  Section~\ref{sec:conclusion} presents our conclusions.

\section{Data} \label{sec:data}

\begin{table*}[ht]
    \centering
    \begin{tabular}{l|c|c|c|c}
       Sample name & Azimuthal slice & $z$ selection & $\#$ & $\#(R>6~\si{kpc})$ \\
       \hline
       Bailer-Jones (BJ)  & $|\phi|<30^{\circ}$ & Basic &  $7,206,369$& $5,387,688$\\
       BJz+  & $|\phi|<30^{\circ}$ & Basic \& $z>0$ & $3,514,109$& $2,723,565$\\
       BJz-  & $|\phi|<30^{\circ}$ & Basic \& $z<0$ & $3,692,260$& $2,664,123$\\
       BJ-30-15  & $-30^{\circ}<\phi<-15^{\circ}$ & Basic & $1,346,210$& $1,023,916$\\
       BJ-150  & $-15^{\circ}<\phi<0^{\circ}$ & Basic & $2,087,847$& $1,579,588$\\
       BJ015  & $0^{\circ}<\phi<15^{\circ}$ & Basic & $2,218,542$& $1,631,103$\\
       BJ1530  & $15^{\circ}<\phi<30^{\circ}$ & Basic & $1,553,770$& $1,153,081$\\
       \hline
       StarHorse (SH)  & $|\phi|<30^{\circ}$ & Basic & $3,731,718$& $2,634,917$\\
       SHz+  & $|\phi|<30^{\circ}$ & Basic \& $z>0$ & $1,736,940$& $1,340,895$\\
       SHz-  & $|\phi|<30^{\circ}$ & Basic \& $z<0$ & $1,994,778$& $1,294,022$\\
       \hline
       Ou+23  & $|\phi|<30^{\circ}$ & Basic & -& $33,335$\\
    \end{tabular}
    \caption{Details of the selections done to make the different samples considered in this work and their number of datapoints.
     Note that the data from  \citetalias{Ou2023} is based on \textit{APOGEE} DR17 combined with photometry from \textit{Gaia} DR3, \textit{WISE} and \textit{2MASS},
     while the BJ and SH samples are based on Gaia DR3 only.
     All of these samples are restricted to only Giants stars.
      \citetalias{Ou2023} select giants with only a cut in $0<\log(g)<2.2$.
     The selection of giants for the BJ and SH samples is stated in the text.
     All these samples have been restricted to our basic spatial cut as described in the text. This involves selecting $|z|<1~\si{kpc}$, but also including stars within $6^{\circ}$ of the Glactic plane to incorporate the flare.}
    \label{tab:selections}
\end{table*}

To model the dynamics of stars in the Milky Way disk reliably, we require accurate distance estimates and velocities.
The large amount of data available in \textit{Gaia} DR3 ($33,812,183$ million stars with radial velocities)
has made it possible to perform model-informed Bayesian analysis to derive distances through informed statistics for a large portion of the catalogue.
In this work, we use two of these \Gaia-based samples to ensure the robustness of our results against bias.

Our primary sample of stars is based on the \Gaia{} RVS catalogue, combined with photo-geometric distances derived by \citet{Bailer-Jones2021}.
This study uses a distance prior and an approximate model of the Galaxy to infer a probabilistic value of the distance to stars in \textit{Gaia} eDR3 that have parallaxes.
We limited ourselves to stars with a relative error in the photogeometric distance smaller than $20\%$. 
We restrict our samples to red giant stars,
effectively removing many faint sources nearby with large uncertainties and reducing completeness problems at large distances.
To identify giant stars, we use stellar parameters estimated by \citet{Andrae2023} to select
$\log(g)<3.0$ and $3000~\si{K}<T_{\mathrm{eff}}<5200~\si{K}$, following \cite{GaiaCollaboration2023Drimmel}.
These criteria leave  $10,497,674$ stars.

We transform the positions and velocities of these stars to Galactocentric coordinates by assuming a solar position
of $R_{\odot}=8.122~\si{kpc}$ and $z_{\odot}=0.025~\si{kpc}$ \citep{distancegalcen,Juric2008},
and a solar velocity of $(11.1,245.6,7.77)~\si{km.s^{-1}}$,
where the $x-$component is taken from \cite{solarmotion} and the $y$ and $z-$components from \cite{Reid2004_galcen}.
We propagate the measurement errors by numerically estimating the Jacobian of the coordinate transformation using the \texttt{vaex} built-in error propagation routine
\citep{Breddels2018}. 

\begin{figure}[t]
    \centering
    \includegraphics[width=\linewidth]{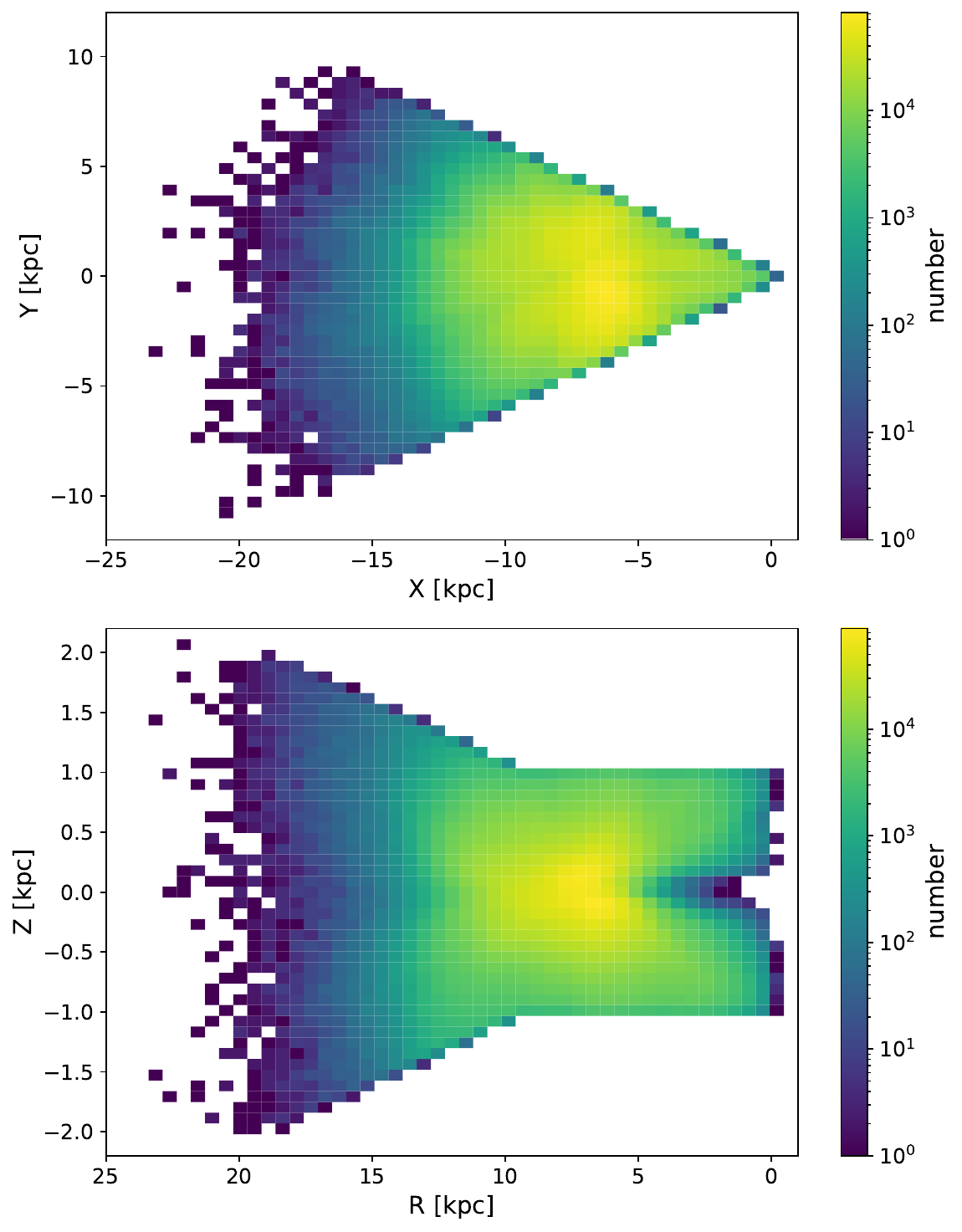}
    \caption{Spatial distribution of the BJ sample after applying the cuts described in Section \ref{sec:data}.
     The top panel shows the $x-y$ plane, whereas the bottom panel shows the $R-z$ plane.  }
    \label{fig:spatialsample}
\end{figure}

Following \citet{Ou2023}, hereafter \citetalias{Ou2023},
we make a series of spatial selections to restrict our samples to the disk plane.
First, we select a wedge in the direction of the Galactic anticentre ($|\phi|<30^{\circ}$),
and to incorporate the flaring of the Milky Way disk, we select $|z|<1~\si{kpc}$,
but for $R>9.5~\si{kpc}$ also include stars within $6^{\circ}$ of the Galactic plane in a wedge in $z$ starting in the Galactic centre.
We apply a cut in $|v_z|<100~\si{km.s^{-1}}$ to remove possible contaminants from the halo.
We refer to this sample as the Bailer-Jones, or the BJ, sample,  containing $7,206,369$ stars,  
whose number density distribution can be seen in Fig. \ref{fig:spatialsample}.
From this sample, we make several subselections to study positionally dependent effects on our results.
These include selections above and below the disk plane and splitting the sample azimuthally.

As a secondary sample,
we use the \Gaia{} RVS combined with distance estimates of the StarHorse (SH) catalogue \citep{Anders2022}.
This study estimates distances in a Bayesian way through isochrone fitting,
using \textit{Gaia} data and photometric data from external catalogues (Pan-STARRS1, SkyMapper, 2MASS, and AllWISE).
We process this dataset similarly to the BJ sample,
making the same cut in relative distance error, 
giant selections, and spatial selections. Addtionally, we clean our sample using the quality cuts of \citet{Antoja2023}.
The final SH sample contains $3,731,718$ stars.
The SH sample is almost completely contained within the BJ sample due to the more restrictive quality cuts of the SH catalogue.
Only $1\%$ of the stars in the SH sample are not in the BJ sample, and these stars are randomly distributed through the Milky Way.

Table~\ref{tab:selections} provides information regarding our samples and spatial subsamples,
and lists also that of \citetalias{Ou2023}. We have a much larger number of stars than  \citetalias{Ou2023} \citep[or][which contains approximately 255,000 stars]{Zhou2023}, but our distance errors are bigger on average and, as a consequence, we only reach out to 20 kpc, compared to 25 - 30 kpc of these other works. 

For the majority of our following analysis, we find similar results between the BJ and SH samples.
For brevity we will focus the majority of our discussions on the primary BJ sample
and place results for the SH sample in Appendix \ref{app:shfigs}.

\section{Jeans Equations in the MW}\label{sec:methods}

The Jeans equations stem from taking velocity moments of the collisionless Boltzmann equation assuming steady-state equilibrium. In the case of axisymmetry the natural coordinates are cylindrical ($R,\phi$, $z$).
The radial Jeans equation, relating to the gravitational force in the radial direction and thus to the circular velocity curve, is \citep[equation 4.222a in][]{Binney2008}:
\begin{equation}
    \dfrac{\partial(\nu\ \langle v_R^2\rangle)}{\partial R} + \dfrac{\partial(\nu\ \langle v_Rv_z\rangle)}{\partial z} + \nu\left(\dfrac{\langle v_R^2\rangle - \langle{v_{\phi}^2\rangle}}{R}+\dfrac{\partial\Phi}{\partial R}\right) = 0,\label{eq:jeanseq}
\end{equation}
where $\Phi(R,z)$ is the Galactic potential, $\nu(R,z)$ is the density distribution of the tracer population and $\langle\cdot\rangle$ signifies the mean of the given quantity.
Equivalent equations can be written down for the vertical and azimuthal components (but the latter is trivially satisfied in the case of axisymmetry as $\partial\Phi/\partial\phi = 0$).
Equation~\ref{eq:jeanseq} reveals that the radial force depends on the second moments of the velocity distributions, their derivatives and the density of the tracers. 

To derive the circular velocity curve from this Jeans equation, we rewrite Equation~\ref{eq:jeanseq} to:
\begin{equation}
\begin{aligned}
    v_{\rm c}^2 = \langle v_{\phi}^2\rangle\ &-\ \langle v_R^2\rangle\left(1+\dfrac{\partial\ln\nu}{\partial\ln R}+\dfrac{\partial\ln\langle v_R^2\rangle}{\partial\ln R}\right)\\
    &-\ R\left(\dfrac{\langle v_Rv_z\rangle}{z}\dfrac{\partial\ln\nu}{\partial\ln z}+\dfrac{\partial\langle v_Rv_z\rangle}{\partial z}\right),\label{eq:vc}
\end{aligned}
\end{equation}
where we have defined that $v_{\rm c}^2(R)=R\left|\dfrac{\partial\Phi}{\partial R}\right|_{z\approx0}$.
The second term in this equation is known as the \textit{asymmetric drift term}, and the third is the \textit{crossterm}.
We model $\nu$ with an analytical function of the form:
\begin{equation}
    \nu(R,z) \propto e^{-R/R_{\mathrm{exp}}-|z|/h_z}\label{eq:numdens},
\end{equation}
where we assume $R_{\mathrm{exp}}=3~\si{kpc}$ and $h_z=0.3~\si{kpc}$ \citep{Ou2023,BlandHawthorn}.
This represents a disk-like number density, which is appropriate since our sample is dominated by stars in the MW disk. In Section \ref{sec:results:systematics} and in Sec.~\ref{sec:discussion} we discuss the effects of using these values (and of other assumptions) on our results. 

To estimate the velocity moments at a given radius,
we bin the stars linearly in $R$ with a width of $0.5~\si{kpc}$ starting at the Galactic centre.
We calculate the errors in our bins through bootstrapping.
To deconvolve results from the measurement errors,
it is necessary  to first calculate $\langle v_R^2\rangle = \langle v_R\rangle^2 + \sigma_{v_R}^2 - \sigma_{\mathrm{obs}}^2$,
where $\sigma_{v_R}^2 $ is the variance of the $v_R$ distribution in a bin,
and $\sigma_{\mathrm{obs}}$ is the propagated error to this quantity from the measurement errors.
We find that the effect of deconvolution with measurement errors is small
since they are typically only $\sim0.01\%$ of our statistical errors when combined with the bootstrapping error.

\subsection{Second moments: $\langle v_R^2\rangle$, $\langle v_{\phi}^2\rangle$ and $\langle v_z^2\rangle$}\label{sec:results:secmoms}

\begin{figure}[th!]
    \centering
    \includegraphics[width=\columnwidth]{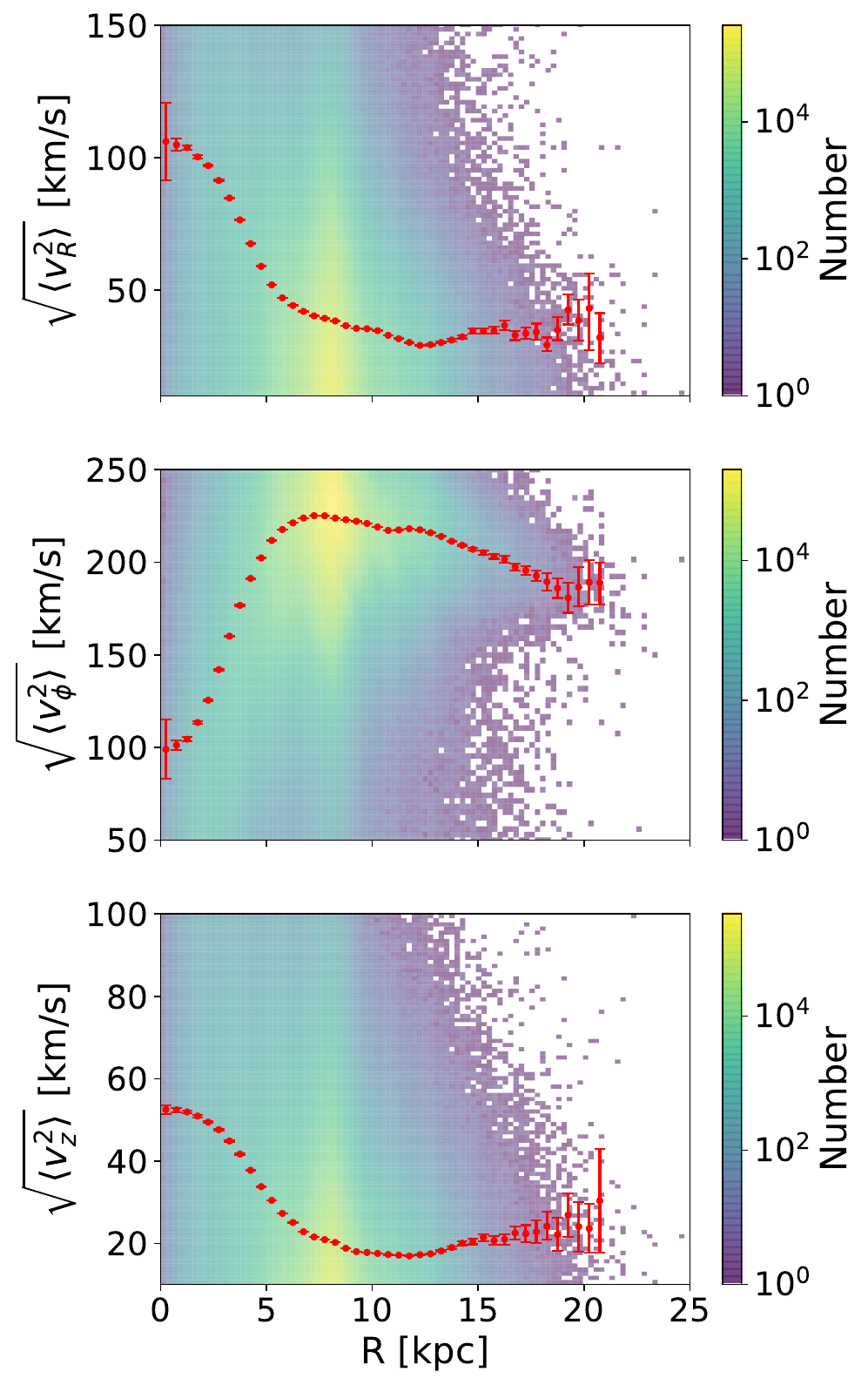}
    \caption{Second moments of the velocity components for the BJ sample.
     From top to bottom the panels show $\sqrt{\langle v_R^2\rangle}, \sqrt{\langle v_{\phi}^2\rangle}$ and $\sqrt{\langle v_z^2\rangle}$.
    The background histograms show the distribution of $||(vv^T-C)_{ii}||$ for our sample:
    the diagonal elements of the velocity tensor for each star.
    The red data points show the square root of the second moment of the sample in the radial bins.
    The error bars show the $1\sigma$ errors from bootstrapping.
    Bins with less than $15$ stars are not shown since they typically produce spurious results.
           }
    \label{fig:radBJ}
\end{figure}

Figure \ref{fig:radBJ} shows the radial profiles and distributions of the BJ sample,
with scattered red points depicting $\sqrt{\langle v_i^2\rangle}$ calculated for each radial bin.
We do not show bins with less than $15$ stars since they typically produce spurious values with large errors.
The background distribution shows the histogram of $\|(vv^T-C)_{ii})\|$ for each star in the sample.
Here, $vv^T$ signifies the velocity tensor for that star, and $C$ signifies the covariance matrix of the observational errors for that star.

For $R<6~\si{kpc}$, the effect of the Galactic bar cannot be neglected,
particularly if it is a long bar as some recent works suggest \citep{Wegg2015,Portail2017,Sormani2022}.
We therefore exclude this region from our analysis as a clear violation of axisymmetry.
From {looking at the background histogram in} Figure \ref{fig:radBJ} we find that $\langle v_{\phi}^2\rangle$ is dominated by the mean $v_{\phi}$, {whereas} the other two velocity moments are dominated by dispersion since the bulk of the stars have
a lower value of the velocity tensor than the radial profile of $\langle v_i^2\rangle$.
There are clear ridges visible in the background histogram for $\langle v_{\phi}^2\rangle$, corresponding to the ridges reported in \citet{Antoja2018}, \citet{Kawata2018} and \citet{Ramos2018}.

\begin{figure}[th!]
    \centering
    \includegraphics[width=\linewidth]{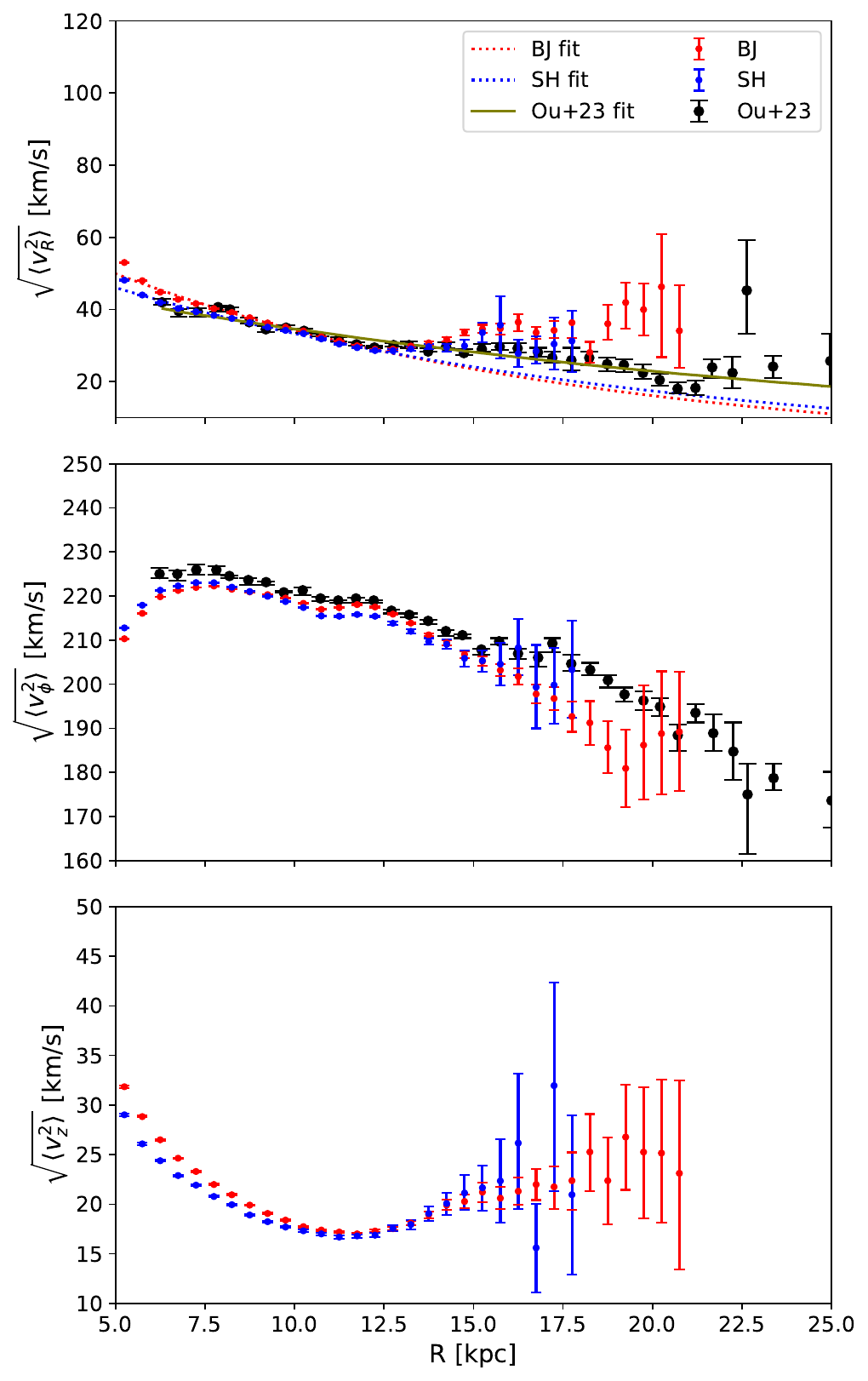}
    \caption{Comparison of the three radial profiles of the second moments in $v_R, v_{\phi}$ and $v_z$ (from top to bottom)
     in the MW stellar disk for the BJ (red) and SH (blue) samples.
     The dotted lines in the top panel are the exponential profile fit from Equation \ref{eq:expfitvR}.
     The error bars show the bootstrapping error from binning.
     The black dots show the radial profiles from Fig. 3 from \citetalias{Ou2023}, and the olive dashed line shows the exponential profile fit to their points.
     The difference in $v_{\phi}$ between our results and
     those from \citetalias{Ou2023} is due to a difference in assumed solar velocity parameters and slight differences in the sample.}
    \label{fig:compBJSH}
\end{figure}
\begin{figure}[th]
    \centering
    \includegraphics[width=\linewidth]{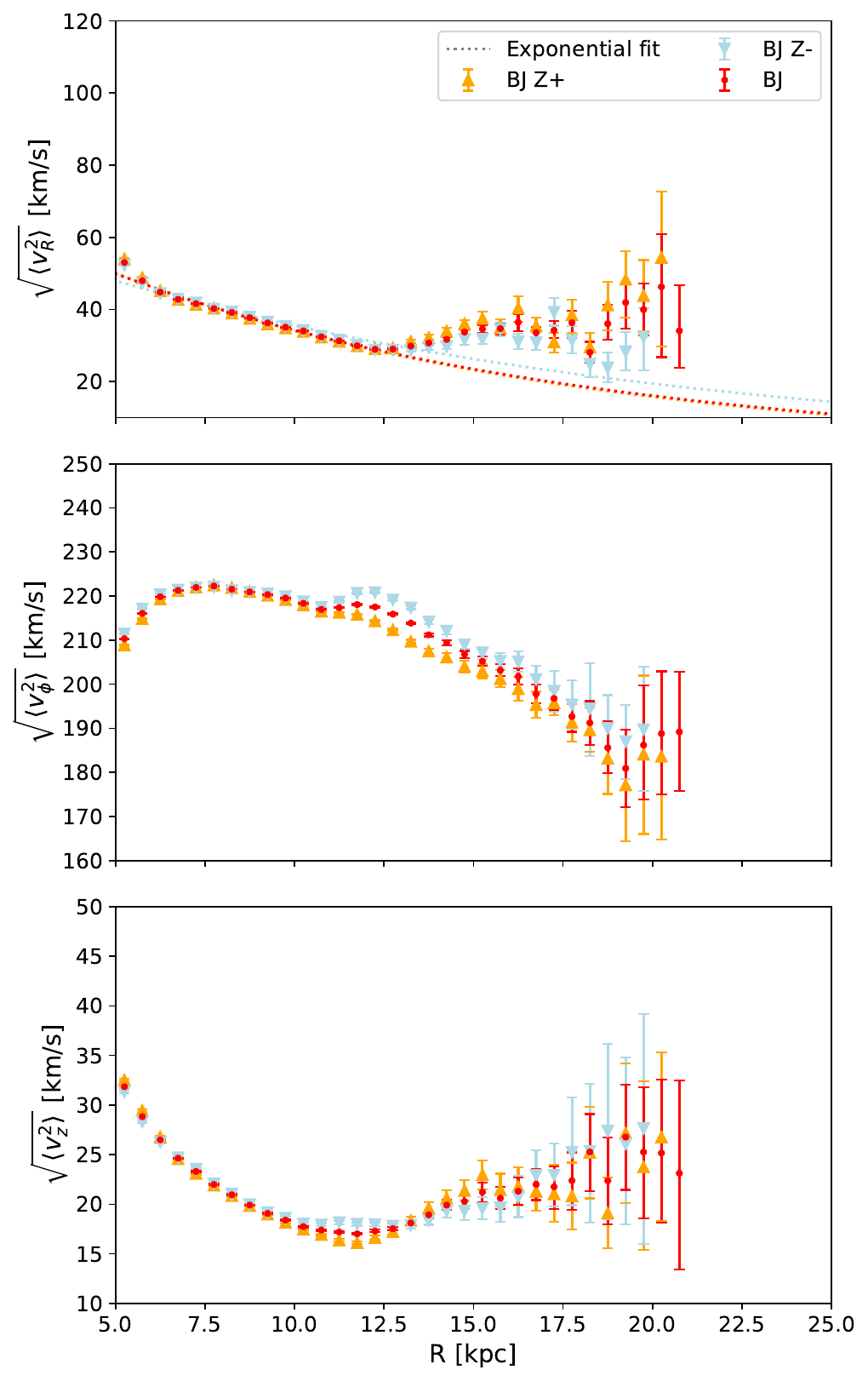}
    \caption{Comparison of the three radial profiles of the second moments of $v_R, v_{\phi}$ and $v_z$ (from top to bottom) in the MW stellar disk for the BJ sample split into a sample above ($z>0$, blue) and below ($z<0$, red) the disk plane.}
    \label{fig:zsplitBJ}
\end{figure}

In Fig.~\ref{fig:compBJSH} we show our inferred radial profiles. The results based on the BJ and SH samples agree within uncertainties for $\langle v_R^2\rangle$ and $\langle v_z^2\rangle$. We have also plotted the  \citetalias{Ou2023} results in this figure.  
Beyond $R\gtrsim17~\si{kpc}$, our bins typically contain less than $\sim50$ stars and thus suffer from low-number statistics. This is where the differences with  \citetalias{Ou2023} are largest. Their smaller distance errors enables them reaching farther out in the disk. The differences for $ < 17$~kpc in the azimuthal second moment (which is systematically lower than that from \citetalias{Ou2023}), can be almost entirely attributed to the difference in solar position and velocity used for coordinate transformations.

{To infer $v_{\rm c}$ using equation \ref{eq:vc} typically a function is fitted to the profile of $\langle v_R^2\rangle$ to avoid numerically noisy derivatives.}
In an {isothermal} disk in dynamical equilibrium with a constant scale-height and mass-to-light ratio, the vertical Jeans equation can be used to show that the second moments of $v_R$ as well as $v_z$ should follow an exponential radial profile {if one assumes an exponential surface density profile} \citep{vanderKruit1981}:
\begin{equation}
    \langle v_R^2\rangle(R) \propto \exp\left(\dfrac{-R}{R_{\mathrm{exp,\langle v_R^2\rangle}}}\right).\label{eq:expfitvR}
\end{equation}
{Under the stated assumptions, $R_{\mathrm{exp},\langle v_R^2\rangle}=2R_{\rm exp}$ should hold.} While this relation has been confirmed to hold in external galaxies, for example by \citet{Martinsson2013}, these authors do report deviations for $R > 1.5 R_{\exp}$. This is perhaps not surprising given that at large distances disks may be flaring and hence breaking down the assumption of a constant scale-height. Nonetheless, following previous works, we fit this profile in a radial range between $6<R<15~\si{kpc}$ (see the top panel of Fig. \ref{fig:compBJSH}).
The typical exponential scale length we find for the selections listed in Table \ref{tab:selections} is $R_{\mathrm{exp},\langle v_R^2\rangle}\approx6.8~\si{kpc}$, which is not too different from the expected $2R_{\rm exp}$.

\begin{figure*}
    \centering
    \includegraphics[width=\linewidth]{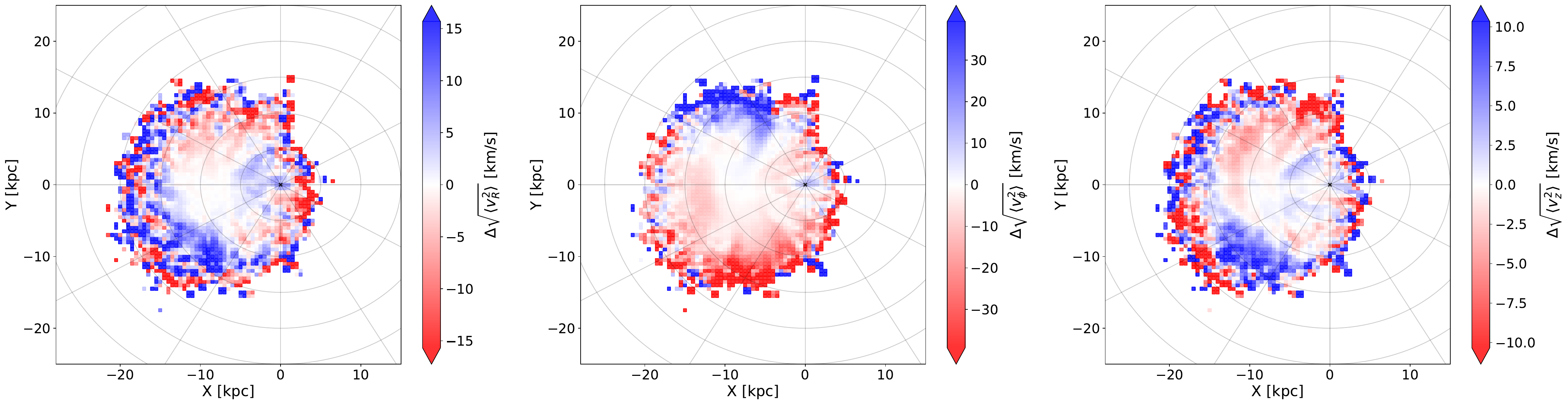}
    \caption{The difference between the distributions of the second moments of $v_R, v_{\phi}$ and $v_z$ (from left to right) for the BJz+ and BJz- samples. We use evenly spaced bins in $x$ and $y$ of width $0.5~\si{kpc}$.
     The colorbars were limited to within the $1\sigma$ value of the distribution of calculated $\Delta\sqrt{\langle v_i^2\rangle}$ for the whole sample.
     }
    \label{fig:zpmBJ}
\end{figure*}

The radial profile of $\langle v_R^2\rangle$ deviates from the fitted exponential profile from $R\approx12.5~\si{kpc}$ onward,
as can be seen also in Fig. \ref{fig:radBJ}.
The deviation from the exponential profile for the second moments of $v_R$ and $v_z$ is less pronounced for the SH sample but is still present.
This means that the magnitude of the deviation is only slightly dependent on the specifics of the sample selection, but its existence is independent of the sample used. 
The flattening of the 2nd moment in $v_R$ is also seen in \citetalias{Ou2023} and \cite{Zhou2023}, but is attributed to a much larger exponential scale-length (indeed the data is fitted out to $\sim 30$~kpc). The work by  \citetalias{Ou2023} find $R_{\rm exp, \sqrt{\langle v_R^2\rangle}}=25~\si{kpc}$, similar to that reported by \citet{Zhou2023}, who report a value of 24~kpc. While such a fit does not work well for our data (it overpredicts the dispersion at $R\sim 10-12$~kpc, and underpredicts the trend at large distances, although this can be due to distance errors), it is indicative of an assumption that does not fully hold in the modelling, likely related to the flaring of the MW disk. 

To study asymmetric behaviour in the MW disk, we divided our samples in various subsets (see Table \ref{tab:selections} for details on these selections).
The first division was made by separating the sample above ($z>0$, BJz+) and below ($z<0$, BJz-) the disk plane. 
Figure \ref{fig:zsplitBJ} shows the radial profiles of the BJz+ and BJz- samples.
The most striking feature is how the profiles of the three velocity components are different for the various subsets.
In $v_R$, the moments above and below the plane start deviating from an exponential profile at $R>12.5~\si{kpc}$, with BJz+ having higher values than BJz-.
In $v_{\phi}$, the split happens at $R>11~\si{kpc}$ with BJz- being above BJz+, in agreement with \citet{GaiaCollaboration2021anti}, see their Fig. 12.
In $v_z$, we see oscillating behaviour, where the different subsamples have crossing points at $R=10,13$ and $16~\si{kpc}$.
So while the disk kinematics are symmetric with respect to the Galactic plane within $R\approx11~\si{kpc}$,
beyond this radius, this is clearly not the case.

We have also inspected the dependence of the second moments on azimuth by splitting the sample in wedges 
of $15^{\circ}$ each. The results are shown in Fig.~\ref{fig:azimBJ} of Appendix~\ref{app:azfigs}, and although
some differences are apparent, they are of smaller amplitude than the asymmetry with respect to the Galactic disk plane.

Fig. \ref{fig:zpmBJ} shows the distribution of the difference between samples BJz+ and BJz- of the three second moments of the velocity in the $x-y$-plane.
Here, we can see clear antisymmetry both azimuthally and vertically throughout the MW disk.
While $v_R$ and $v_z$ show depict differences of up to $10~\si{km.s^{-1}}$ for $R < 10$~kpc,  
$v_{\phi}$ shows more asymmetry, up to $20~\si{km.s^{-1}}$.

\begin{figure}[t]
    \centering
    \includegraphics[width=\linewidth]{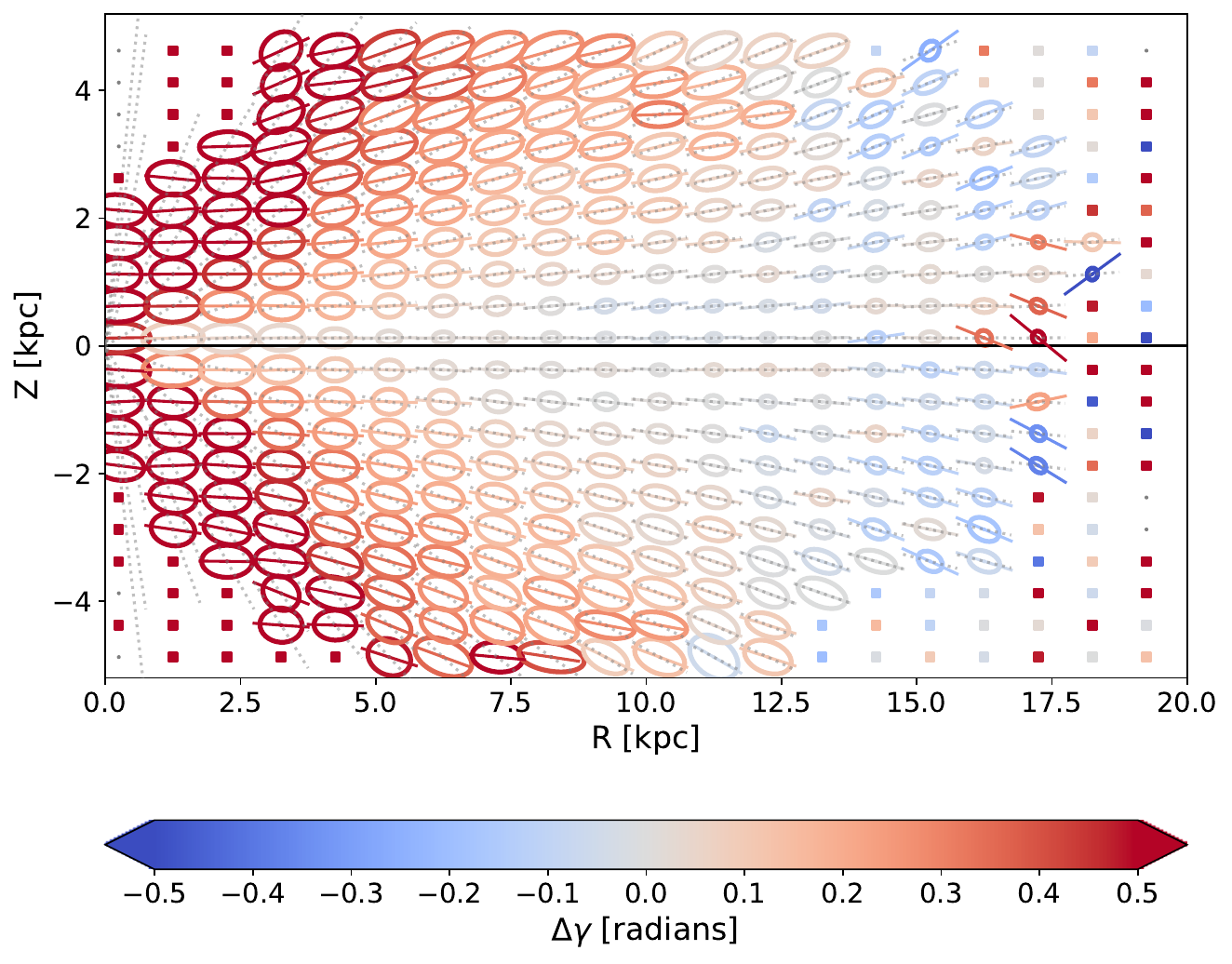}
    \caption{The misalignment of the velocity ellipsoid with respect to spherical alignment in radians, {plotted in cylindrical coordinates $R,z$}.
     For $z>0$ this is defined as $\gamma-\gamma_{\mathrm{sph}}$ and for $z<0$ as $\gamma_{\mathrm{sph}}-\gamma$.
      Square markers denote bins with less than $25$ stars.
       We only plotted every other bin of width $0.25~\si{kpc}$ in $R$ and $z$ for clarity of the figure.
        }
    \label{fig:tiltBJ}
\end{figure}

\subsection{The Correlation $\langle v_Rv_z\rangle$ and Crossterm} \label{sec:results:crossterm}

The crossterm (the final term in Eq.~\ref{eq:vc}) is typically neglected in calculations of the circular velocity curve
\citep[see e.g.][who estimate the contribution to $v_c$ to be smaller than a few percent]{Eilers2019,Ou2023,Jiao2023}.
To estimate its magnitude for our dataset, we first find the correlation $\langle v_Rv_z\rangle(R,z)$,
using additional bins in $z$ with a width of $0.25~\si{kpc}$.
To calculate the value of $\langle v_Rv_z\rangle$ at greater heights above and below the disk plane
we relax the cut in vertical coordinates to $|z|<5\kpc$.
The errorbars are estimated by bootstrapping the sample in each bin.

The behaviour of $\langle v_Rv_z\rangle$ throughout the MW disk can be quantified by the tilt angle $\gamma$ of the velocity ellipsoid:
\begin{equation}
    \tan(2\gamma) = \dfrac{2\langle v_Rv_z\rangle}{\sigma_{v_R}^2-\sigma_{v_z}^2}.\label{eq:tiltangle}
\end{equation}

The tilt angle of the stars in the BJ sample, with the relaxed $z$ cut, can be seen in Fig.~ \ref{fig:tiltBJ}.
For every bin, we plot an ellipse representing the velocity ellipsoid, except for bins with less than $25$ stars, which are given by a square marker.
If the tilt of the velocity ellipsoid follows a spherical alignment, this reduces to $\gamma_{\mathrm{sph}}=\tan^{-1}(R/z)$.
We define the mismatch between $\gamma$ and $\gamma_{\mathrm{sph}}$ to be $\gamma-\gamma_{\mathrm{sph}}$ for $z<0$ and $\gamma_{\mathrm{sph}}-\gamma$ for $z>0$. This mismatch with respect to spherical alignment is given by the color coding in Fig.~\ref{fig:tiltBJ}. 
We find that the velocity ellipsoids seem to be more aligned with the plane of the stellar disk than with a spherical coordinate system,
in agreement with \cite{Hagen2019}.
However, there is also some asymmetry in the $z=0$ plane for $|z|>1.5~\si{kpc}$.

\begin{figure}[t]
    \centering
    \includegraphics[width=\linewidth]{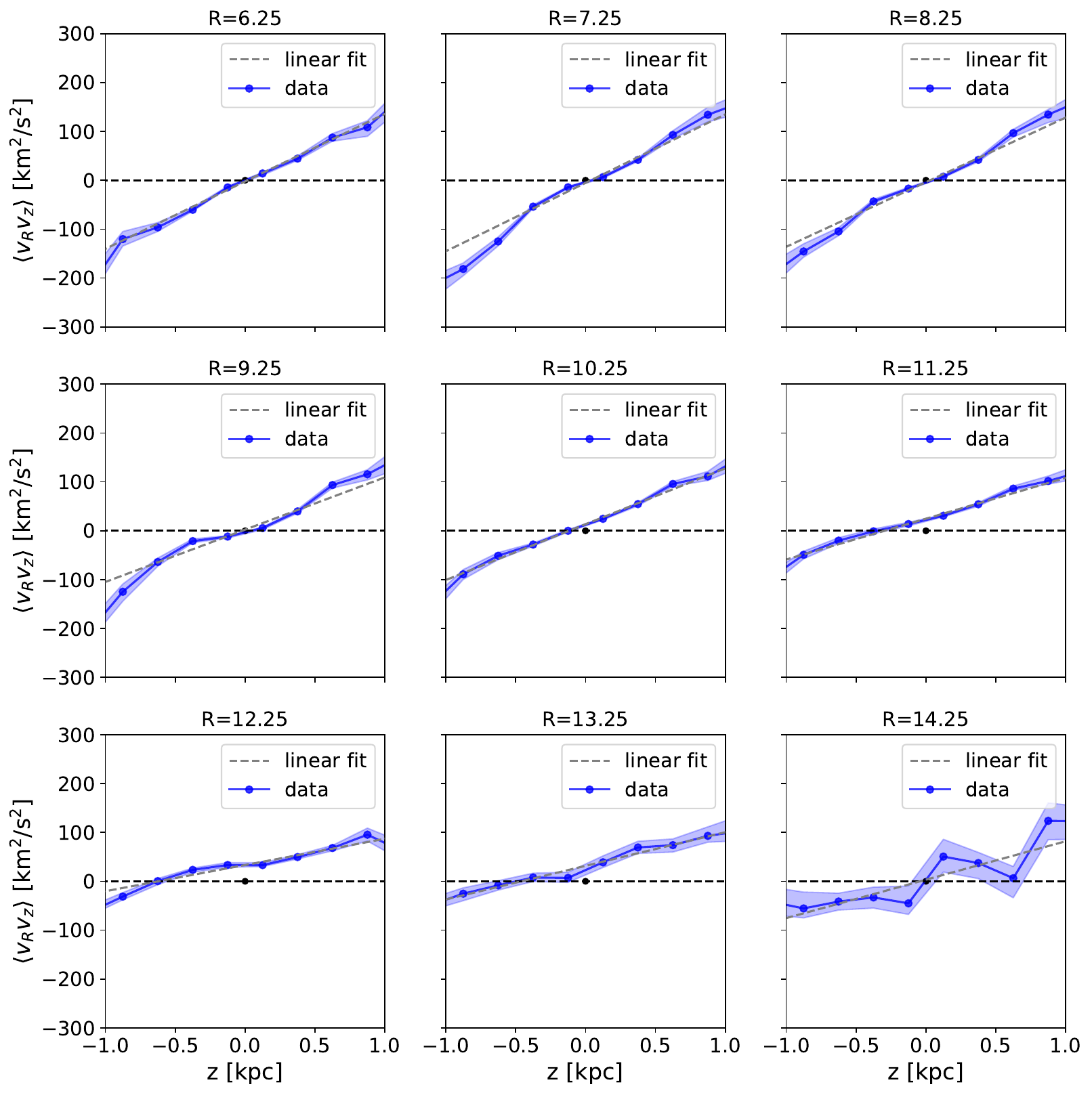}
    \caption{Vertical profile of the correlation $\langle v_Rv_z\rangle$ for different bins in $R$.
     The shaded region shows the bootstrapped error from binning.
      The dashed grey line is a linear fit to the data between $|z|<1$.
       The dashed black line can guide the eye to where the correlation crosses the origin.
        The origin is denoted with a black dot.
        }
    \label{fig:crosstermBJ}
\end{figure}

For calculating the influence of the crossterm in $v_{\rm c}$ as function of radius,
 we proceeded to fit a profile to $\langle v_Rv_z\rangle$ to estimate $\partial\langle v_Rv_z\rangle/\partial z$ in Equation \ref{eq:vc}.
We chose to fit a linear profile here, such that we can write the crossterm as:
\begin{equation}
    R\left(\dfrac{\langle v_Rv_z\rangle}{z}\dfrac{\partial\ln\nu}{\partial\ln z}+\dfrac{\partial\langle v_Rv_z\rangle}{\partial z}\right) = R(Az+C)\left(\dfrac{A}{Az+C}-\dfrac{\text{sign}(z)}{h_z}\right),\label{eq:crossterm}
\end{equation}
where $\langle v_Rv_z\rangle (z)=Az+ C$. 
In the fully axisymmetric case, $\langle v_Rv_z\rangle$ is expected to follow a linear profile that passes through the origin ($C$ = 0).
To estimate the slope $A$, 
we fit the data for $|z|<1~\si{kpc}$ using eight bins (see Fig.~\ref{fig:crosstermBJ}).
We find that a linear fit to the correlation $\langle v_Rv_z\rangle$ is still reasonable, although the data does not necessarily cross the origin. Previous work by \citet{Silverwood2018} fitted $\langle v_Rv_z\rangle(z)=Az^n$ for only $z>0$, a 
profile that also assumes that the correlation crosses the origin,
and hence cannot capture the asymmetry above and below the plane seen in Fig.~\ref{fig:crosstermBJ}.

The assumption of a linear behaviour seems to break down for $|z|>0.5~\si{kpc}$ for $R=7.25$~kpc,
and we see a slight deviation from linear behaviour at $8.25$ and $9.25~\si{kpc}$ as well.
Even though the linearity seems to hold within $|z|<0.75~\si{kpc}$ for $R>10~\si{kpc}$ again
(although for $R=14.25~\si{kpc}$ non-linear behaviour may be apparent, it is had to assess given the low number of stars),
this is where a clear offset of $\langle v_Rv_z\rangle$ from crossing the origin starts.
This offset is very likely due to the warp present in the MW disk,
whose the mean $z$-coordinate is indeed negative for $R>10~\si{kpc}$ and for the Galactic longitudes probed by our sample {(see e.g. \citet{Skowron2019,Cabrera-Gadea2024} or \citet{Jonsson2024} for a first analysis of the effect of the warp on the local dynamics).}
These findings imply the need to be careful in interpreting results from the axisymmetric Jeans equations at $z=0$ as the plane of symmetry no longer coincides with $z=0$. 

\section{The Velocity Curve of the Milky Way and its uncertainties}\label{sec:rotation curve}
\begin{figure*}[ht]
    \centering
    \includegraphics[width=0.85\linewidth]{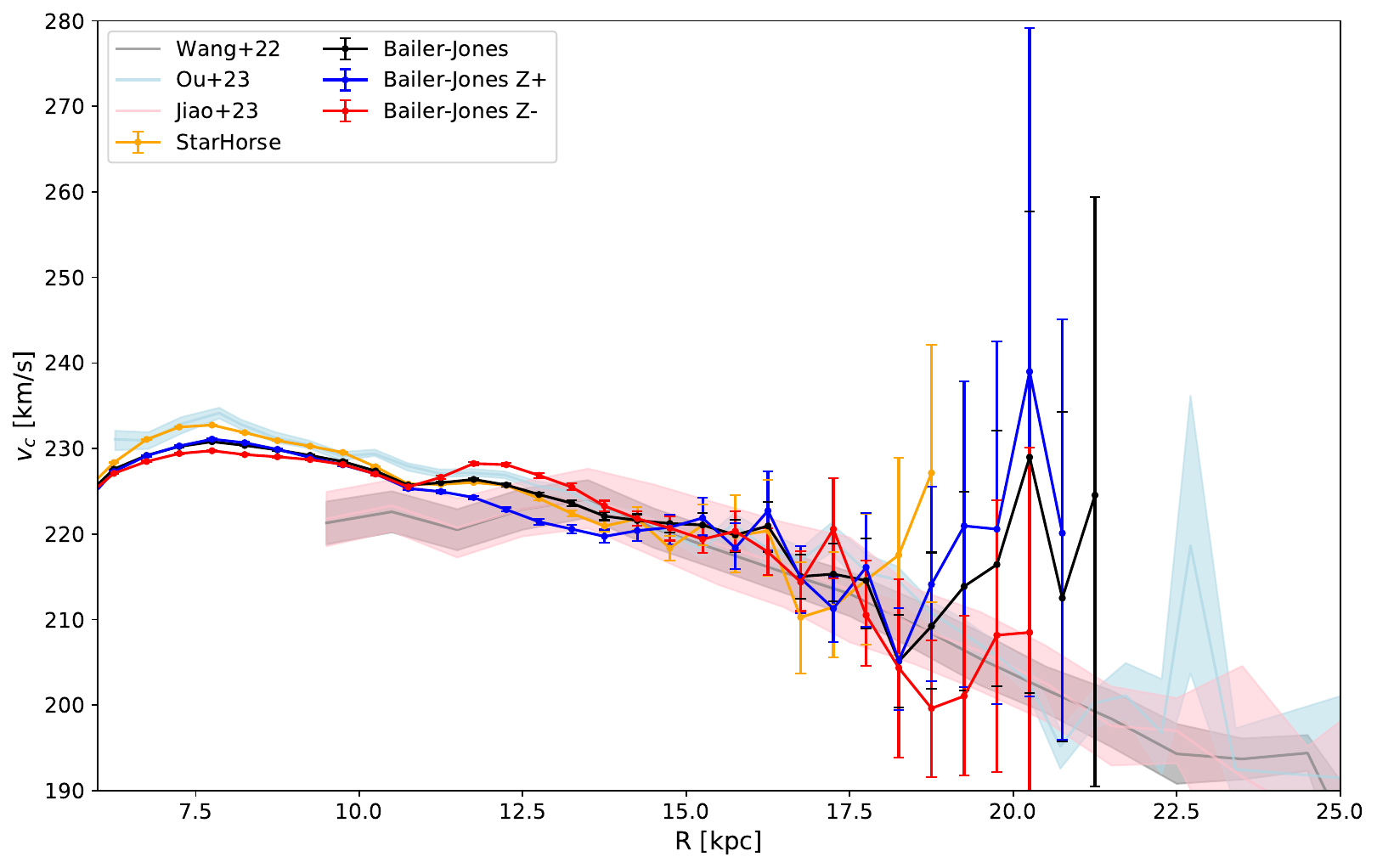}
    \caption{Rotation curves calculated for several samples from Table \ref{tab:selections}.
     The black and {orange} lines show the BJ and SH samples, respectively.
      The blue and red lines show the BJz+ and BJz- samples, respectively.
       The grey, lightblue, and pink lines and shaded regions show the results from \citet{Wang2023}, \citetalias{Ou2023}, and \citet{Jiao2023}, respectively.
       }
    \label{fig:compvc}
\end{figure*}

\subsection{Derivation of the circular velocity curve}
By combining the individual terms of Equation~\ref{eq:vc}, we calculate the circular velocity curve for our samples and present it in Fig. \ref{fig:compvc}. This figure reveals a decline in our inferred circular velocity curve that starts at $R\sim 7.5$~kpc and continues at least out to $R\sim 15$~kpc. This is in good agreement with the previous studies of \citet{Wang2023}, \citet{Zhou2023}, \citet{Ou2023} and \citet{Jiao2023} and that are shown as shaded regions, 
within the statistical uncertainties of these studies for $R>12~\si{kpc}$. For $R<12~\si{kpc}$, the difference in our results from those from  \citetalias{Ou2023} is mostly caused, as discussed earlier, by a difference in assumed solar velocity. For large distances, $R>17.5~\si{kpc}$, our derivation of the circular velocity curve starts to suffer from the effects of low number statistics.  

Splitting our sample above and below the disk plane reveals a statistically significant difference of $\sim 8\kms$ in the circular velocity curve using data located above or below the plane. This difference is a direct consequence of the split observed in the middle panel of Fig. \ref{fig:zpmBJ}, which is driven by the behaviour of $\langle v_{\phi}^2\rangle$, which is asymmetric with respect to the Milky Way disk plane. This is likely due to the Galactic warp. While at $R>18~\si{kpc}$ asymmetries also are apparent, this is also where our uncertainties sharply increase, and we are unable to assess whether the circular velocity curve follows a more steep decline as reported in \citet{Wang2023,Ou2023,Zhou2023}. 

\begin{figure*}[h]
    \centering
    \includegraphics[width=0.85\linewidth]{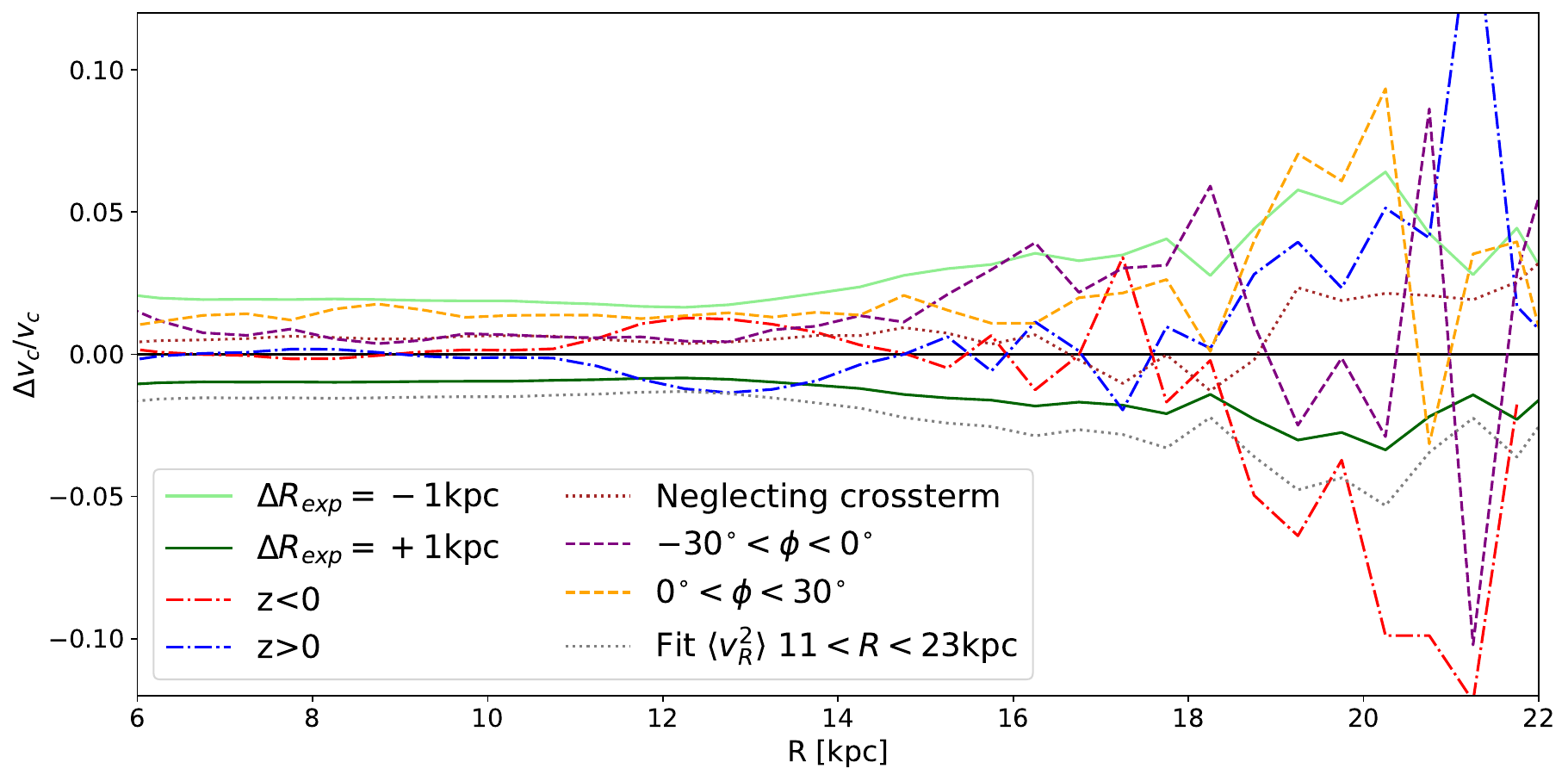}
    \caption{Systematics when using Jeans equations to calculate $v_{\rm c}$.
     Lines show the difference between different selections or methods and the fiducial BJ model.
     The lightgreen (darkgreen) solid line shows the systematic shift when changing the exponential scale length of the number density
     by $-1~\si{kpc}$ ($1~\si{kpc}$).
     The red (blue) dash-dotted line shows the variation expected considering samples below (above) the disk, selecting $z<0$ $(z>0)$ only.
     The {red} dotted line shows the influence of the crossterm in Equation \ref{eq:crossterm}.
     The dashed purple and orange lines show $\Delta v_{\rm c}/v_{\rm c}$ for samples for which $-30<\phi<0^{\circ}$ and $0<\phi<30^{\circ}$ respectively.
     The dotted grey line shows the influence of trying to fit $\langle v_R^2\rangle$ to its behaviour in the outskirts, resulting in $R_{\rm exp,\langle v_R^2\rangle}=43~\si{kpc}$. The black line guides the eye to zero difference with the fiducial model.}
    \label{fig:systematicsBJ}
\end{figure*}

\subsection{Assessment of the systematic uncertainties}\label{sec:results:systematics}

The systematic uncertainties caused by our various assumptions are summarised in Fig.~\ref{fig:systematicsBJ}, which is 
similar to figures presented in previous studies such as the earlier cited \citet{Eilers2019},\citet{Zhou2023},\citet{Ou2023} and \citet{Jiao2023}.
Individually, under $R\approx18\kpc$ the systematic effects discussed here are below $5\%$, which is of order $11.5~\si{km.s^{-1}}$.

\begin{itemize}
\item \textit{Number Density Profile}: 
The selection function of our survey is arguably too poorly understood to recover an accurate number density of the tracer population,
forcing us to assume a given analytic profile.
We choose exponential decline (Eq.~\ref{eq:numdens}) with a scale length of $R_{\rm exp}=3~\si{kpc}$ following previous works.
The light green (dark green) solid lines in Fig. \ref{fig:systematicsBJ} show the systematic shift expected for the calculated $v_{\rm c}$ when
we lower (raise) $R_{\rm exp}$ by $1~\si{kpc}$.
This can have an effect up to $5\%$ in the outer regions and to $2\%$ within $R<14~\si{kpc}$.
The size of the effect of raising $R_{\rm exp}$ is smaller than that of lowering it by the same amount,
and thus summarizing this uncertainty into one line is a misrepresentation.
Alternatively, \citetalias{Ou2023} investigated the systematic differences found by modelling the number density as a power law instead,
finding a greater systematic uncertainty beyond $10\kpc$. We return to this point in Sec.~\ref{sec:discussion}.

\item \textit{Radial Velocity Profile}: 
To compute a smooth derivative of the radial velocity profile
$\langle v_R^2\rangle$ we assume and fit an exponential profile (Eq.~\ref{eq:expfitvR}).
However, this does not describe the data well for $R>12.5~\si{kpc}$ (see Fig. \ref{fig:compBJSH}).
If the data is instead fitted in the region of $11<R<22~\si{kpc}$, we obtain a much larger scale-length of 43~kpc,
which reflects the flattening off seen at large radii in $\langle v_R^2\rangle$ for our dataset.
The grey dotted line in Fig. \ref{fig:systematicsBJ} shows the difference found from using the larger scalelength on the rotation curve,
which amounts to $2.5\%$ for $R<13~\si{kpc}$, and up to $5\%$ at large radii. 
    
\item \textit{Asymmetries and Spatial Selection}: 
The Jeans equations we apply assume a time-independent, axisymmetric potential.
As shown through vertical or azimuthal cuts in Sec.~\ref{sec:results:secmoms}, this assumption is not true throughout the disk.
The red (blue) dash-dotted lines in Fig. \ref{fig:systematicsBJ} show the effect of only looking at the sample below (above) the disk,
whilst the purple (orange) dashed lines show the effect of looking only at the wedge in $-30<\phi<0~^{\circ}$ ($0<\phi<30~^{\circ}$).
Each of these selections has an effect that can be as large as $3\%$.

Additionally, we find less restrictive spatial selections have approximately a $1-2\%$ effect on our results.
Extending the cut in $|z|$ from $<1~\si{kpc}$ to $<5~\si{kpc}$ has only an effect of about $0.5\%$ on the results for $v_{\rm c}$.
Relaxing the additional $6^{\circ}$ wedge in $z$, chosen to ensure the dataset includes the flaring of the disk, impacts results by less than $1\%$.
These results suggest that our results do not depend strongly on the specific choice of spatial selection.

\item \textit{Crossterm}: The contribution of the crossterm in Equation \ref{eq:crossterm} is shown as the brown dotted line, and its 
contribution 
is under $0.5\%$ for $R\leq17~\si{kpc}$ and can be neglected when compared to other systematic effects.

\item \textit{Local Standard of Rest}: 
A difference in the value of the local standard of rest velocity can cause a systematic shift in $\langle v_{\phi}^2\rangle$,
correspondingly of up to $3\%$ in $v_{\rm c}$ at any given radius (Fig. \ref{fig:compvc}).
This estimate is slightly larger than that by \citetalias{Ou2023},
who separately tested the effect of changing the Sun's distance to the Galactic centre and the proper motion of SgrA* within their errors.
This shift (visible in Fig.~\ref{fig:compBJSH}) is the likely cause of the differences between our results and those of \citetalias{Ou2023} for $R<12\kpc$.
\end{itemize}

\noindent We note that different selections or systematics can shift the resulting $v_{\rm c}$ at each radius either up or down,
and combining these effects into a single error estimate is not trivial because they are not independent or Gaussian in nature \citep[see e.g.][for a discussion of the impact of this assumption]{Oman2024}.
In a worst-case scenario, assuming that they are independent, they could add up in quadrature to give an error of $\sim6\%$ for $R<14~\si{kpc}$
and up to $15\%$ at larger radii.
This is in accordance with the findings from \citetalias{Ou2023}, \citet{Jiao2023} and typically larger than estimated by \cite{Zhou2023}.

This is particularly relevant in the context of recent attempts to determine the density profile of the Galaxy's dark matter halo
by fitting the circular velocity curve \citep[as done in e.g.][]{Ou2023,Jiao2023}.
These works favour halos with a steep outer slope or cutoff, which is driven by a declining rotation curve for $R>15~\si{kpc}$. The uncertainties in our dataset prevent us from seeing such a steep decline beyond this radius. In any case, 
this region is where the systematics are higher, and it is good to recall that a 15\% systematic on the circular velocity 
results in a 30\% uncertainty on the mass enclosed at these radii.

\section{N-Body simulations of an interaction with a heavy perturber}\label{sec:simulations:Sagittarius}
To investigate the effectiveness of the radial Jeans equation in recovering the circular velocity curve of a system,
we apply our Jeans methodology to two simulation suites of  MW-like galaxies.
This also allows us to study possible causes of the asymmetry we found in data from \textit{Gaia} DR3 
and their likely impact on the derived \vc ~for the MW. In this section we focus on the Simulation L2 from \cite{Laporte2018},
an N-body simulation of a MW-like galaxy interacting with a satellite system like the Sagittarius dwarf.
\subsection{Description of the simulation L2}

The Sagittarius progenitor in the L2 simulation has an initial mass of $M_{200}=6.0\times10^{10}M_{\odot}$.
This relatively large mass leaves a remnant of Sagittarius that is arguably too massive at present day compared to observations.
Nonetheless, it successfully reproduces qualitatively important observables like the phase space spiral,
and the Monoceros and TriAnd structures in \textit{Gaia} DR2 \citep{Laporte2019}.
The initial conditions of the disk stars were generated to reproduce in broad terms the Milky Way. For example, the density profile described by an exponential disk with a scale length of
$R_{\mathrm{exp}}=3.5~\si{kpc}$, which is also what we use in our application of Jeans equation. {For each snapshot, the disk was aligned by iteratively finding the center of mass, and then aligning the $z-$axis to the angular momentum of the inner disk \citep[as done in e.g.][]{Laporte2018,Garcia-Conde2024}. We can thus expect similar complications from a warp present in the disk of L2 as we saw for the MW (see Sec. \ref{sec:results:crossterm}).}

\begin{table}
    \centering
    \begin{tabular}{c|c|l}
        Name & Time $[\si{Gyr}]$ & Note \\
        \hline
        $t_1$ &  $1.9$ & Dynamically relaxed disk\\
        $t_{\mathrm{peri},1}$ &  $2.4$ & 1\textsuperscript{st} pericenter\\
        $t_{2}$ &  $2.6$ & Just after 1\textsuperscript{st} pericenter\\
        $t_{3}$ &  $4$ & Just before 2\textsuperscript{nd} pericenter\\
        $t_{\mathrm{peri},2}$ &  $4.5$ & 2\textsuperscript{nd} pericenter\\
        $t_{4}$ &  $4.6$ & Just after 2\textsuperscript{nd} pericenter\\
        $t_{\mathrm{peri},3}$ &  $5.6$ & 3\textsuperscript{rd} pericenter\\
        $t_{5}$ &  $5.7$ & Just after 3\textsuperscript{rd} pericenter\\
        $t_{\mathrm{peri},4}$ &  $6.1$ & 4\textsuperscript{th} pericenter\\
        $t_{\mathrm{peri},5}$ &  $6.4$ & 5\textsuperscript{th} pericenter\\
        $t_{6}$ &  $6.9$ & Present day analogue\\
    \end{tabular}
    \caption{Timestamps of interest in the L2 simulation from \cite{Laporte2018} that we consider in this work. $t_{\mathrm{peri},i}$ is listed to show when the pericenters of the satellite occur. The $t_i$ given are the snapshots we consider for our analysis in this Section.}
    \label{tab:laportetimes}
\end{table}

We select simulation samples using the disk particles in two $60^{\circ}$ wedges on opposite sides of the disk,
denoted as $x<0$ and $x>0$. 
We repeat our spatial selection from Section \ref{sec:data}, with a cut in $|z|<0.5~\si{kpc}$ instead of $1~\si{kpc}$.
This change affects our results by less than $1\%$. 

We specifically choose $6$ snapshots in this simulation, corresponding to just before or after a pericenter passage of the satellite
(see Table \ref{tab:laportetimes}).
These pericenters are the moments when the satellite has the strongest dynamical effects on the disk
and demonstrate the impact on the Jeans modelling the clearest.
The first two pericenters of the satellite are at $\sim30~\si{kpc}$ and $\sim20~\si{kpc}$ respectively,
while the other pericenters are within $\sim15~\si{kpc}$.
The mass of the satellite is still significant at the later $3$ passages while interacting more directly with the disk system.

The true velocity curve of the simulation can be calculated as 
$v_{\rm c, true}=\sqrt{-a_RR}$, with $a_R$ being the acceleration in the $R$-direction due to the gravitational force enacted.
We exclude particles belonging to the satellite,
which typically have negligible effect outside of the satellites pericenter,
and they are not a part of the system we want to apply Jeans equations to.
We estimate the intrinsic variation of $v_{\rm c, true}$ around the disk by calculating $v_{\rm c, true}$ in wedges of 15 degrees around the disk and calculating the dispersion at each radius.
\begin{figure*}[ht]
    \centering
    \includegraphics[width=\linewidth]{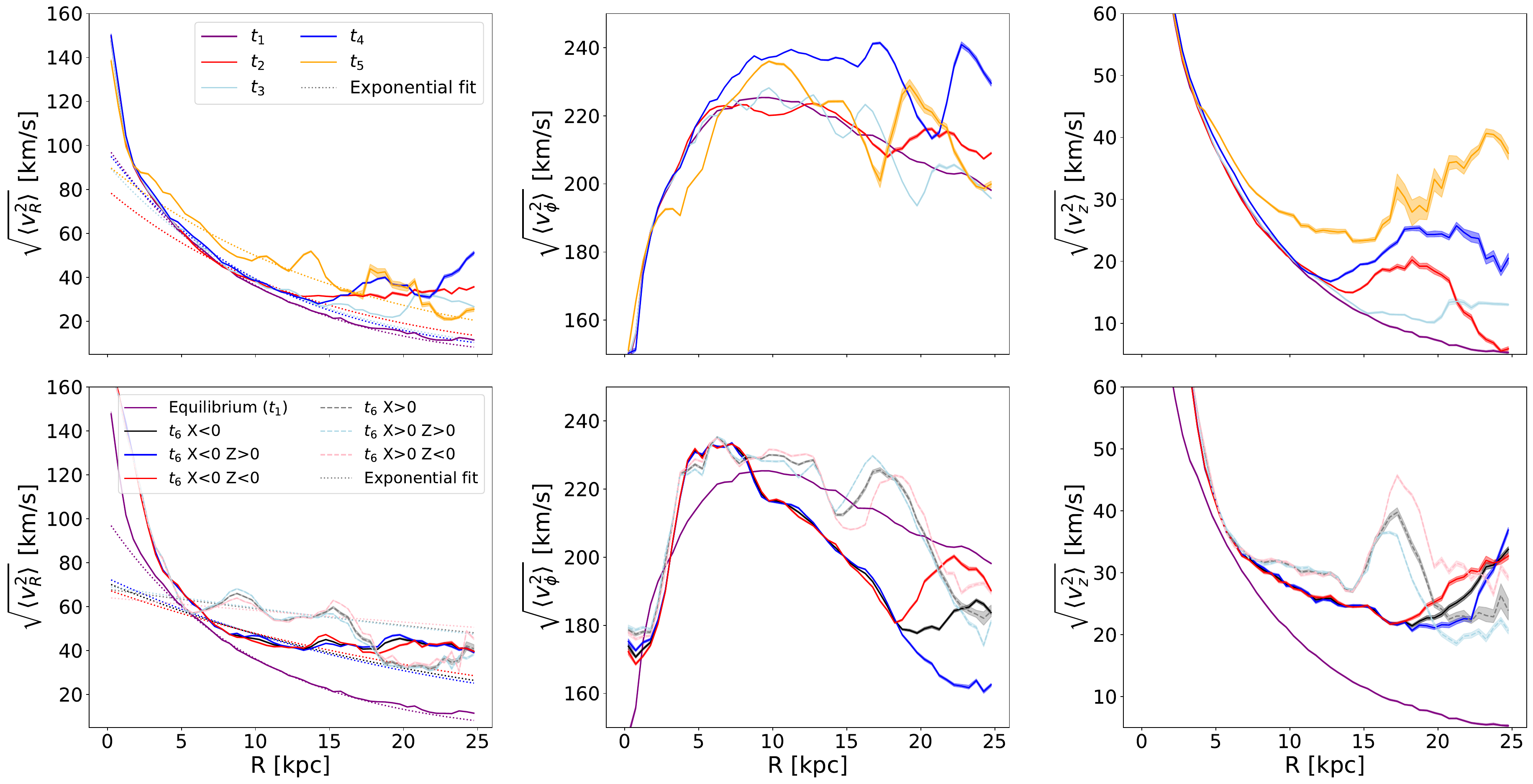}
    \caption{Radial profiles of the three second moments of the velocity distribution in simulation L2 from \cite{Laporte2018}. The top row shows $t_{\{1,2,3,4,5\}}$. The bottom row shows $t_{6}$ for both sides of the disk in $x\gtrless0$ and vertical cuts in $z\gtrless0$, along with the equilibrium situation at $t_1$ for reference.}
    \label{fig:Laporteradialprofs}
\end{figure*}
\begin{figure*}[ht]
    \centering
    \includegraphics[width=\linewidth]{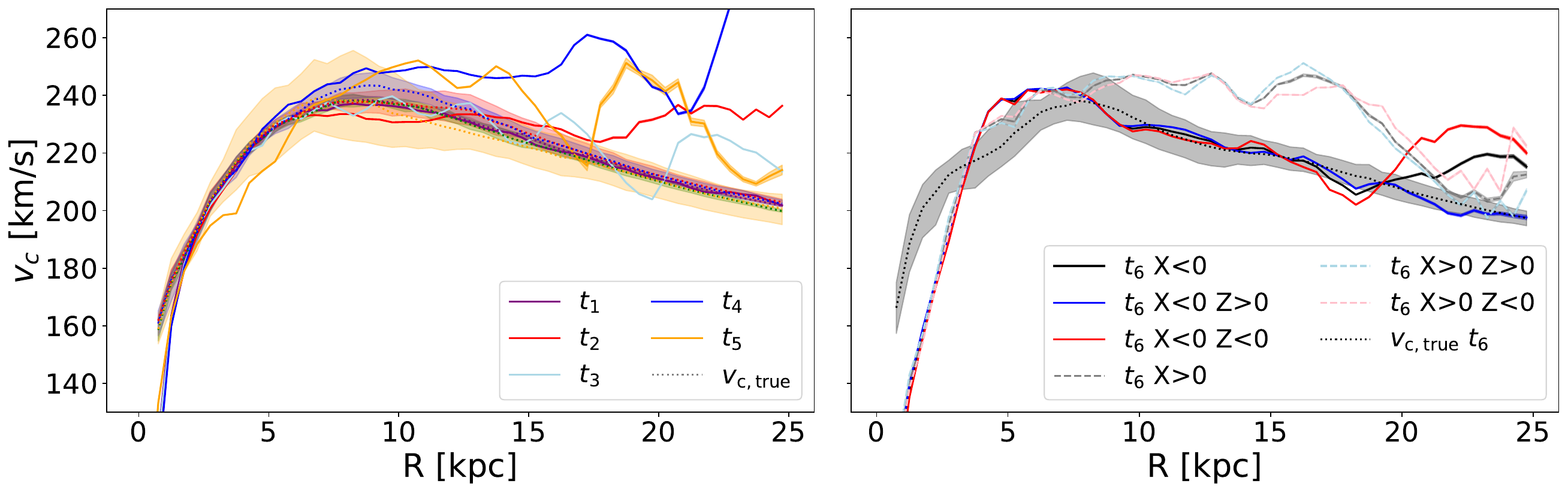}
    \caption{$v_{\rm c}$ calculated for the different snapshots in simulation L2 from \cite{Laporte2018}. The {dotted} lines show {the azimuthally averaged value of} $v_{\rm c, true}$, with the shaded region showing the $1\sigma$ quantiles of the intrinsic variation. The left panel shows $t_{\{1,2,3,4,5\}}$, corresponding to the top row of Fig. \ref{fig:Laporteradialprofs}. The right panel corresponds to the bottom row of Fig. \ref{fig:Laporteradialprofs}, showing the results for $t_6$ in different slices for $x\gtrless0$ and $z\gtrless0$.}
    \label{fig:Laporteradialvcs}
\end{figure*}

\subsection{Applying Jeans Equations to L2}
The radial profiles of the second velocity moments, calculated for the snapshots at times $t_i$ from Table \ref{tab:laportetimes},
can be seen in Fig.~\ref{fig:Laporteradialprofs}, which can be compared to Fig. \ref{fig:compBJSH}. The top row shows results for $x<0$ for all snapshots except $t_6$, which we show in the bottom row with additional cuts in $z\gtrless0$ and $x\gtrless0$.
Initially (at $t_1$) the disk is dynamically relaxed and symmetric and $\langle v_R^2\rangle$ and $\langle v_z^2\rangle$ follow an exponential decline.
At later times, we see an increase in these second moments in the outskirts of the disk. This effect is larger just after an interaction ($t_{2,4,5,6}$),
and then as the disk relaxes the effect decreases ($t_3$), although it never vanishes at large radii. 

The second moment for $v_{\phi}$ depicts wiggles which reflect the formation of spiral arms in the disk as can be seen from the {lightblue} curve for $t_3$.
The first interaction indeed produces tidally induced spiral arms as seen in Figs. 14 and 15 from \cite{Antoja2022} using the same simulation.
Furthermore, we can see the effect of the longer dynamical timescale in the outer disk region.
While the second moment of $v_R$ follows the exponential profile for $R<20~\si{kpc}$ for $t_3$, it shows a deviation for $R>20~\si{kpc}$,
similar to the deviation for $R>12.5~\si{kpc}$ just after the first pericenter at $t_2$.

At $t_4$ (after the second pericenter) and $t_5$, the dispersions of the velocities are high,
and we see how the interactions move the disk further toward disequilibrium and asymmetries.
For $t_6$ we see clear differences in the recovered second moments for $x>0$ and $x<0$, and also a split in the profiles above and below the plane,
even though the overall behaviour is similar.
The split is less pronounced for $v_R$ and $v_z$ than it is for $v_{\phi}$.
This split is likely related to bulk motions arising in disk stars in the outer disk due to the interactions with the satellite.
We found a similar split  in the MW data (see Fig. \ref{fig:zsplitBJ}), where there is a difference up to $10~\si{km.s^{-1}}$ at $R\approx13~\si{kpc}$.
In contrast, in this simulation the magnitude can be as high as $40~\si{km.s^{-1}}$ as seen for $x<0$ at $R\approx22~\si{kpc}$.
This is potentially due to the interactions with a much heavier Sagittarius,
as its larger mass could be the driver of the much larger asymmetry seen than measured for the MW.

Figure \ref{fig:Laporteradialvcs} shows the $v_{\rm c}$ calculated for our six snapshots {along with the azimuthally averaged value of $v_{\rm c,true}$ and its $1\sigma$ quantiles}.
The left panel shows the results for all snapshots but $t_6$, and $t_6$ itself is plotted in the right panel for the same spatial selections as in the bottom row of
Fig. \ref{fig:Laporteradialprofs}.
The departures from the initial equilibrium configuration (at $t_1$) that were seen in Fig. \ref{fig:Laporteradialprofs} at higher radii are imprinted in the circular velocity curve as well for all other snapshots.
These include the wiggles corresponding to the spiral arm structure at $t_3$, $t_4$, $t_5$ and $t_6$ and 
the splits in the profiles for $t_6$ with $z\gtrless0$.
We see that a higher deviation from a dynamically relaxed and symmetric system (i.e. seen in snapshot $t_1$) is also reflected in a larger mismatch with $v_{\rm c,true}$.

\begin{figure}[ht]
    \centering
    \includegraphics[width=\linewidth]{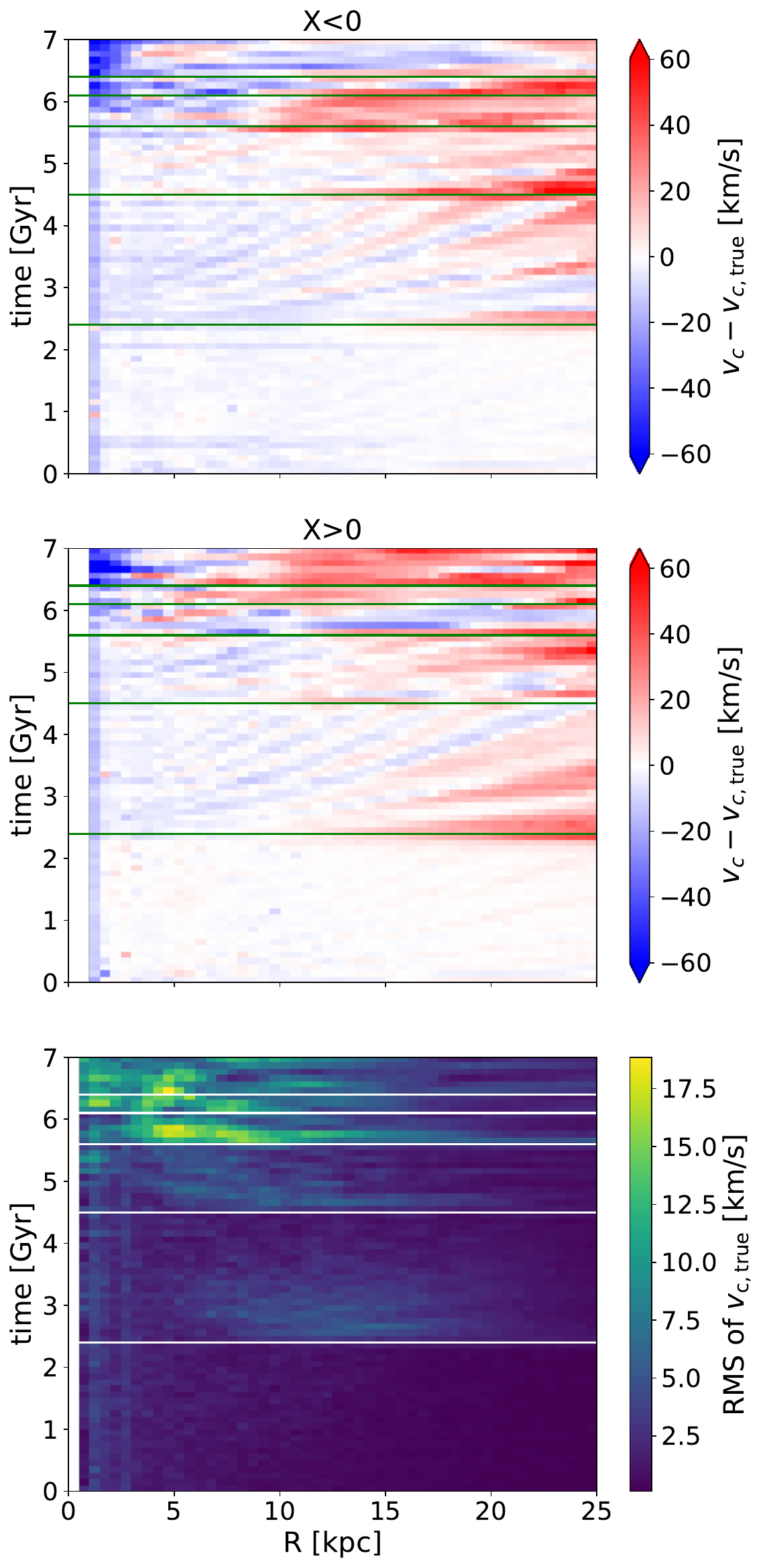}
    \caption{Time-evolution of $v_{\rm c}$ in the simulation from \cite{Laporte2018}. \textit{Top and Middle:} the difference between the Jeans equation derived $v_{\rm c}$ and the true $v_{\rm c, true}$. The top panel shows the result for $x<0$ and the middle panel shows the opposite side of the disk ($x>0$). \textit{Bottom:} the RMS of $v_{\rm c, true}$ with azimuth. Green (white) horizontal lines show the pericenters of the satellite. The start of the simulation is at $t=0$.}
    \label{fig:Laportetimeplots}
\end{figure}

The time evolution of the difference $\Delta v_{\rm c} =v_{\rm c}- v_{\rm c,true}$ can be seen in Fig.~\ref{fig:Laportetimeplots}
for the samples $x>0$ and $x<0$ located on opposite sides of the disk throughout the time of the simulation. {We now only averaged $v_{\rm c, true}$ over the azimuthal wedge under consideration}.
Between the first pericenters,
the $\Delta v_{\rm c}$ shows a structure of diagonal stripes that are the result of the tidally induced bi-symmetric spiral arms,
alternating between areas of overestimating and underestimating the \vc.
On average, there is a tendency to overestimate $v_{\rm c}$ at $R>15~\si{kpc}$ in both sides of the disk,
and to underestimate it at inner radii.
We also looked at the side of the disk at a $90$ degree angle ($y>0$) and found that the sign of $\Delta v_{\rm c}$ is opposite to the slices at $x\gtrless0$
but they also show an overall trend of overestimating (underestimating) at the outer (inner) disk regions at later times.

For later pericenters, the density structures are more complex and do not show perfect 2-fold symmetry.
Hence, the circular velocity is underestimated for $x<0$ (top panel) and overestimated for $x>0$ (middle panel).
During the third pericenter, the satellite more directly interacts with the disk system while still being very massive,
which is reflected in the larger values for $\Delta v_{\rm c}$.

The bottom panel shows the intrinsic variation with azimuth of $v_{\mathrm{c,true}}$. 
The RMS of $v_{\rm c, true}$ at $t_1$ is negligible, while at $t_6$ this intrinsic variation covers a range of $\sim20~\si{km.s^{-1}}$.
The RMS increases slightly between $R=10-15~\si{kpc}$ after each pericenter, and the formation of a bar is apparent after the third pericenter.
The dispersion in the outer regions of the disk remains steady for most of the time.
We have also found in the snapshots that the disk plane is not consistently at $z=0$ after the third pericenter, especially in the outer regions.
This further increases the over- and underestimations in $v_{\rm c}$ seen in the top and middle panels of Fig. \ref{fig:Laportetimeplots} over the whole range in $R$ 
as the disk is highly perturbed around these times.
The shift of mean $z$ determined for snapshots after the third pericenter can reach values of $|\langle z\rangle|=1.5~\si{kpc}$ in the outer regions of the disk. {\cite{Laporte2018} already showed that the presence of Sgr causes a torque on the disk both through the wake in the halo and its own interactions, which are the likely causes of this misalignment.} We conclude from Fig. \ref{fig:Laportetimeplots} that the Jeans equations can not recover $\bar{v}_{\mathrm{c,true}}$ within $5\%$
for $R>8~\si{kpc}$ from the third pericenter onward, and that this mismatch is typically larger (up to $15\%$) for higher radii, and can be either an over- or underestimation.

In summary, we find similar behaviour in the L2 simulation as that resulting from the analysis of the BJ sample,
namely both a flattening off of $\langle v_R\rangle$ and $\langle v_z\rangle$ and an asymmetry in $\langle v_{\phi}^2\rangle$ with respect to the disk plane. 
However, we do not expect the deviations from the true $v_{\rm c,true}$ in the MW to be as big as those in this idealized simulation with a satellite
that is likely more massive than the Sagittarius dwarf galaxy \citep{Laporte2019,Antoja2018}. 

\section{Cosmological simulations} \label{sec:simulations:auriga}
\begin{table*}[]
    \centering
    \begin{tabular}{c|c|m{1.5cm}|m{1.5cm}|m{1.5cm}|c|c|c|m{1.5cm}|m{1.5cm}}
        Name & $R_{\rm exp}$ (kpc) & thin disk $h_{z,8\mathrm{kpc}}$ (kpc) & thick disk $h_{z,8\mathrm{kpc}}$ (kpc) & thin disk $h_{z,18\mathrm{kpc}}$ (kpc) & $R_{\rm flare}$ (kpc) & $z_{\rm cut}$ (kpc) & $\alpha$ ($^{\circ}$) & $N_{\rm stars}\ G<14$ +QC+ errors mock & $N_{\rm stellar particles}$ Simulation \\
        \hline
        MW & $3.0$ (a) & $0.26$ (b) & $0.75$ (b) & $0.64$ (b) & $9.8$ (b) & $1.0$ (c) & $6$ (c) & $7,206,369$ & --\\
        \hline
        Au6 & $3.5$ & $0.33$ & $1.0$ & $0.60$ & $9.10$ & $1.0$ & $6.28$ & $5,879,204$ & $1,359,810$ \\
        Au16 & $7.2$ & $0.30$ & $1.2$ & $0.52$ & $14.3$ & $0.91$ & $3.65$ & $5,530,273$ & $1,109,807$ \\
        Au21 & $4.1$ & $0.50$ & $1.2$ & $1.10$ & $9.10$ & $1.5$ & $9.50$ & $11,662,340$ &$1,624,646$ \\
        Au23 & $5.7$ & $0.34$ & $1.1$ & $0.55$ & $14.3$ & $1.0$ & $4.10$ & $7,976,781$ & $1,623,935$ \\
        Au24 & $6.9$ & $0.33$ & $1.1$ & $0.61$ & $14.3$ & $1.0$ & $4.0$ & $4,787,465$ & $1,276,577$ \\
        Au27 & $3.8$ & $0.30$ & $1.0$ & $0.40$ & $25.0$ & $0.9$ & $2.10$ & $10,551,221$ &$1,757,772$ \\
         
    \end{tabular}
    \caption{Disk parameters for the MW and Auriga galaxies.
     Values for the MW were taken from \cite{BlandHawthorn}, \cite{Chrobakova2022} and \citetalias{Ou2023} (denoted by a, b, c respectively).
     Values for scale heights and flare parameter of the Auriga disks were taken from \cite{Grand2017}.
     The scale length was fitted to the simulation snapshot, and agree with the scale length found in \cite{Grand2017} for Au6, Au16 and Au24.
     The value for $z_{\rm cut}$ is taken to be $3h_{z,8\mathrm{kpc}}$ for the Auriga disks.
     The angle $\alpha$ for the cut in $z$ is calculated such that $\alpha=\tan^{-1}(z_{\rm cut}/R_{\rm flare})$.
     $R_{\rm flare}$ is found by fitting $h_{z}(R)\propto \exp(R/R_{\rm flare})$.
     The number of stars in our observed mock sample is shown in the second to last column.
     The rightmost column shows the number of simulation particles in the sample after performing the spatial selection on the particles in the snapshot.}
    \label{tab:aurigaia}
\end{table*}

To gain even more insight in the applicability of Jeans equations to infer $v_{\rm c}$,
we also apply our methodology to the six simulated galaxies from the \texttt{Auriga} project \citep{Grand2017},
and their corresponding \textit{Gaia} mocks known as \texttt{Aurigaia}  \citep{Grand2018aurigaia}.
This project is a suite of high-resolution magneto-hydrodynamical zoom-in cosmological simulations of MW-sized galaxies \citep{Grand2017},
originally comprising of $30$ simulations labeled Au$1$ to Au$30$ at resolution level 4.
Six of the original $30$ haloes (Au$6$, Au$16$, Au$21$, Au$23$, Au$24$, Au$27$) were then resimulated at a higher resolution (level 3) with
a dark matter particle mass of $m_{\mathrm{DM}}=4\times10^4~M_{\odot}$,
a stellar particle mass of $m_{\mathrm{b}}=6\times10^3~M_{\odot}$ and a softening parameter $\epsilon=184~\si{pc}$.
The properties of the galaxies can be found in Table~\ref{tab:aurigaia}.
Of the 6 haloes,
Au6 is considered a close analogue of the MW considering its stellar mass, star formation rate and morphology.
Au16 and Au24 have larger disks than the MW, and Au21, Au23 and Au27 have had some combination of
heavy, recent and frequent satellite interactions \citep{Grand2018aurigaia}. {For these analyses, the coordinate system has been chosen such that the $Z$-axis is the eigenvector of the inertia tensor of the particles within $0.1R_{200}$ that is most aligned with the largest angular momentum vector of the same particles in the original simulation frame \citep{Grand2017}.}

For these 6 high-resolution simulations, the \texttt{Auriga} project created a mock \textit{Gaia} DR2-like catalogue (see \citet{Grand2018aurigaia} for details).
First, each star particle was expanded into full stellar populations,
drawn from isochrones defined by the star particles' metallicity and age using a Chabrier Initial Mass Function \citep[IMF][]{chabrier03GalacticStellarSubstellar}.
The stars' positions and velocities are then distributed around the approximate phase space volume of the parent star particle by a 6D smoothing kernel,
as described in \citet{Lowing2015}.
These phase-space positions correspond to the ``true'' positions and velocities of the stars,
which can then be transformed to ``observed" Galactic coordinates when choosing a fixed position for the Sun.
From the absolute values of the stars magnitude and colours, the apparent values are calculated, and an additional effect of dust extinction is applied.
Only stars with $G<20$ are included in the \texttt{Aurigaia} mock samples,
simulating the \textit{Gaia} DR2 observability limit.
The original \texttt{Aurigaia} catalogues were created using approximates for the expected \textit{Gaia} DR2 errors using the apparent magnitudes, colours and distances of the stars.
For this work, we recalculate the mock observational errors to resemble \textit{Gaia} DR3-like errors using the \textsc{PyGaia}\footnote{https://github.com/agabrown/PyGaia}
software package.
The observed DR3 mock values of the star are then the true values convolved with these errors.

\begin{figure*}
    \centering
    \includegraphics[width=\linewidth]{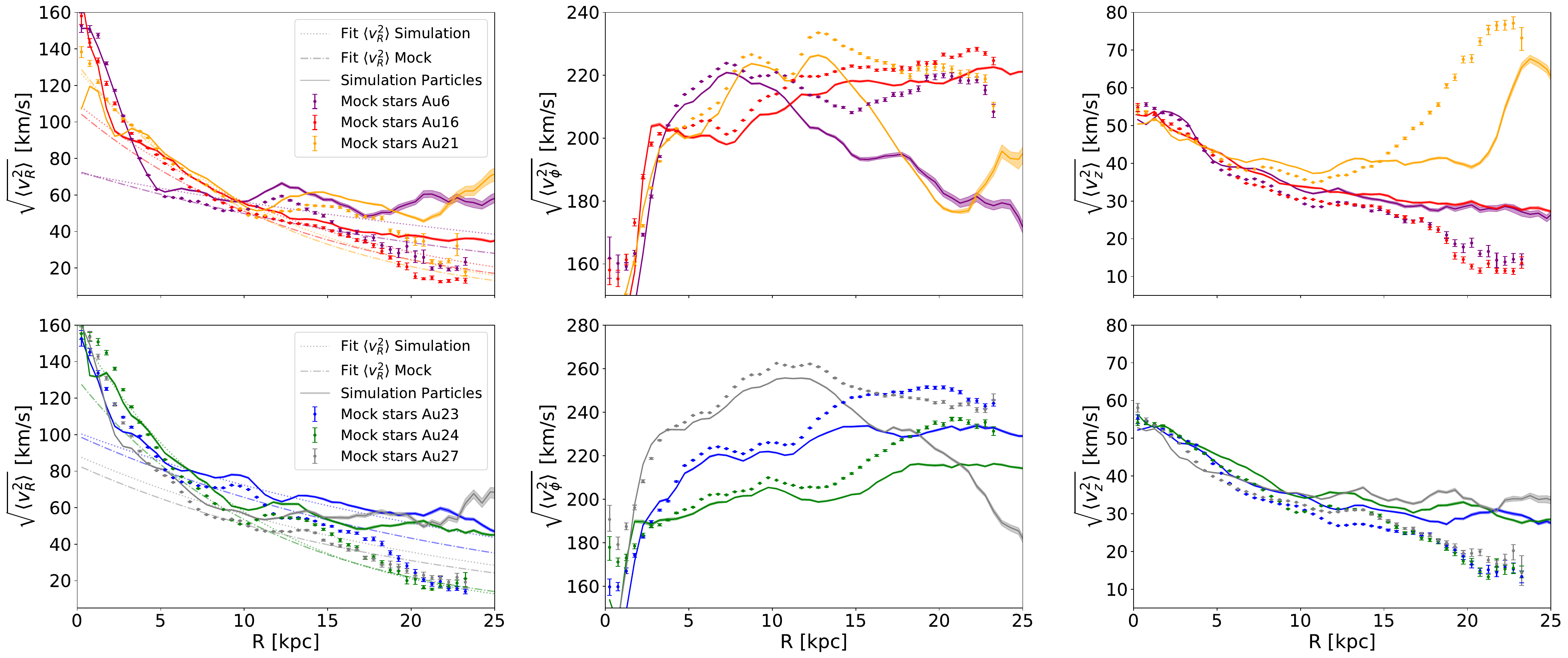}
    \caption{Radial profiles of the second moments of the velocity for the six Aurigaia mocks from \cite{Grand2018aurigaia}.
     The top row has results for Au6, Au16 and Au21, the bottom row shows Au23, Au24 and Au27.
     Scattered points with errorbars show the results for the mock samples described in Table \ref{tab:aurigaia}
     while solid lines show the result for the stellar particles in the final simulation snapshot of the simulated galaxies.
      The dashed lines in the left column show the exponential fit to the profile of the mock sample,
       while the dotted lines show the fit to the profile of the simulation particle sample.
       }
    \label{fig:aurigaiahalos}
\end{figure*}

\begin{figure*}[t]
    \centering
    \includegraphics[width=\linewidth]{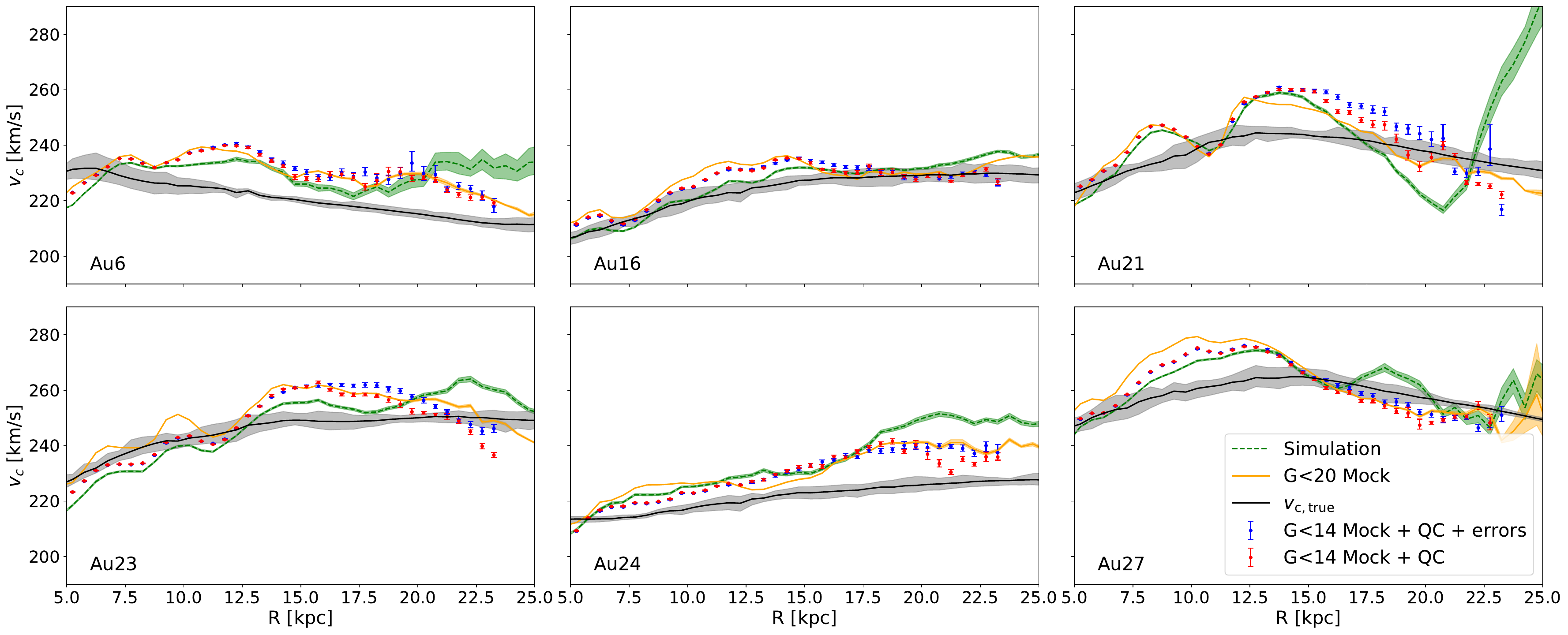}
    \caption{Jeans derived $v_{\rm c}$ for the six Aurigaia mocks from \cite{Grand2018aurigaia}.
     The green dashed line with shaded region shows the result for the simulation particles.
     The orange line shows the result for the full mock catalogue with only a magnitude limit at $G<20$.
     The red points show the results for the full mock with our data quality cuts.
     The blue points show the result for the observed mock sample.
     This is the sample shown in Fig. \ref{fig:aurigaiahalos} and described in Table \ref{tab:aurigaia}.
     The black line and shaded region show $v_{\rm c,true}$ and its associated $1\sigma$ variation.}
    \label{fig:aurigaiadiff}
\end{figure*}

We therefore have three different stellar datasets for every \Auriga{} halo;
the original simulation star particles,
the true stars without any observational errors,
and the DR3 mock stars.
Accompanying this, 
the \Auriga{} project supplies force grids of the final snapshots of the six galaxies,
 which we can use to find $v_{\mathrm{c,true}}$ in the simulation {as we did in Sec. \ref{sec:simulations:Sagittarius} by azimuthally averaging over the force grid}.
The force found in this way is consistent with the force found by directly averaging the gravitational force stored for all simulation particles for the provided snapshot
in the same spatial selection as described in Section \ref{sec:data}. 

For all samples, we perform a similar spatial cut to our MW sample described in Section \ref{sec:data}.
The cut in $|z|<z_{\rm cut}$ and the angle $\alpha$ of the wedge in $z$ is adapted to incorporate the thickness and flare (if present)
of the disk of the simulated galaxy.
This flare is characterised by $R_{\rm flare}$, which is found by fitting $h_z(R)\propto \exp(R/R_{\rm flare})$ \citep{Grand2017}.
The cut $|z|<z_{\rm cut}$ is taken to be three times the scale height at solar radius, $3h_{z,8\mathrm{kpc}}$, so that $\alpha=\tan^{-1}(z_{\rm cut}/R_{\rm flare})$,
which reflects the spatial cut made for the data but scaled to the disk of the simulations.
We also remove stars with $|v_z|>100~\si{km.s^{-1}}$.

To simulate a \textit{Gaia} DR3 RVS selection, we select stars with $G<14$.
For the mock stars, we demand additional quality cuts following our MW selection in Sec.~\ref{sec:methods};
requiring a relative error in parallax smaller than $20\%$,
and making the same  red giant stars selections in $T_{\rm eff}$ and $\log(g)$.
Distances are estimated by inverting the parallax,
and the transformation from ``observed'' to Galactic coordinates is made with the ``true'' solar parameters used to generate the mock. Furthermore, we require the simulation particles that the mock stars were generated from also follow our spatial cuts.

\subsection{Applying Jeans Equations to Auriga}
Figure \ref{fig:aurigaiahalos} shows the radial profiles of the velocity moments for the observed mock samples (symbols with error bars)
and the stellar particles (solid lines with shaded regions) in the final snapshot of the simulation for the $6$ \Auriga{} galaxies.
We see that $\langle v_R^2\rangle$ for the mock sample does not follow an exponential decline from $R\approx10~\si{kpc}$ onward,
rather showing a similar behaviour to the \textit{Gaia} DR3 data and the simulation of the interaction of a Sagittarius-like satellite from Sec.~\ref{sec:simulations:Sagittarius}.
Following our approach to the MW, 
we fit the exponential profile  to $\langle v_R^2\rangle$ in the radial range where the profile is a good description of the data.

The second moment in $v_{\phi}$ of the mock sample shows considerable features indicating deviations from equilibrium and axisymmetry in all galaxies,
for instance the wiggles indicating spiral arms in Au21 ({orange} curve).
For $\langle v_z^2\rangle$ we see similar deviations from an exponential decline as for $\langle v_R^2 \rangle $, but due to the lower dispersions the variations are less pronounced here.
For the sample of mock stars, we typically see a less pronounced departure from an exponential decline for both $v_R$ and $v_z$ compared to the simulation particles,
as well as less distinct features in $v_{\phi}$ pointing to disequilibrium.
Especially for the higher radii this is partially due to the contamination from halo star particles,
since there is no specific selection of the simulation particles, apart from the requirement that $|v_z|<100~\si{km.s^{-1}}$. 

We have also studied the vertical asymmetry in the \texttt{Auriga} systems by again splitting up the sample in a cut above and below the disk. We find a similar behaviour in the profiles to that seen in the data in Fig. \ref{fig:zsplitBJ}. The difference between the samples above and below the disk is of a similar magnitude to the data as well, and smaller than seen in the simulation in Section \ref{sec:simulations:Sagittarius} (Fig. \ref{fig:Laporteradialprofs}). Au6 is most similar in the magnitude of this asymmetry when compared to the MW.

Figure \ref{fig:aurigaiadiff} shows different estimations of the circular velocity curve for all six galaxies. The black line and corresponding grey-shaded region show $v_{\rm c,true}$ and its intrinsic variation with azimuth throughout the disk, following our aproach in Sec.~\ref{sec:simulations:Sagittarius}. We typically see a spread of $\sim1.5\%$ around the azimuthally averaged profile.  The green dashed line shows $v_{\rm c}$ calculated from applying the Jeans Eq. to the simulation particles in the final snapshot, using the radial profiles shown with solid lines in Fig. \ref{fig:aurigaiahalos}.

\subsection{The effects of sample selection}
We now consider the differences between the results obtained for the simulation particles, true stars, and mock observed stars.
To estimate the effect of the RVS selection function on our results we consider the full stellar catalogue;
similar to the true star catalogue but relaxing the RVS magnitude limit of $G<14$ to the $G<20$ magnitude limit of the \texttt{Aurigaia} catalogues.
Additionaly, we require the parent simulation particles of the star also satisfy our spatial selection.
The orange line in Fig. \ref{fig:aurigaiadiff} shows the result for this full stellar catalogue.
Compared to the green line it shows the additional effect of having a selection function applied to the tracer population sample.
We also study the effect of our quality cuts.
By applying these to the full mock sample we end up with the red points in Fig. \ref{fig:aurigaiadiff}. 
Lastly we can see the effect of the measurement errors by looking at the observed mock sample.
This is shown as the blue points in Fig. \ref{fig:aurigaiadiff}, and corresponds to the symbols with error bars in Fig. \ref{fig:aurigaiahalos}.

From these results we see that the simulation particles (green dashed lines) typically provide the closest match to $v_{\rm c,true}$
(black line in Fig. \ref{fig:aurigaiadiff}), with approximately a $5\%$ over- or underestimation on average.
We also see that the simulation particle results tend to start overestimating $v_{\rm c,true}$ for $R\gtrsim20~\si{kpc}$,
which is due to the fact that the disk ends around that radius, so our sample is starting to be dominated by stellar halo particles.
The wiggles in the simulation particle result for Au21 and Au23 at $R<15~\si{kpc}$ are due to spiral arm structure in the disk \citep{Gomez2017}. 

We typically see a $5-10\%$ mismatch between $v_{\rm c}$ and $v_{\rm c,true}$ for the full mock catalogue (orange lines in Fig. \ref{fig:aurigaiadiff}),
which is slightly larger than the result for the simulation particles themselves. 
We investigate how much of this effect is due to the procedure of generating the mock stars from the simulation particles in Appendix \ref{app:mocks}.
We conclude that this effect is typically of the order of $2-3~\si{km.s^{-1}}$ in $v_{\phi}$ and $v_R$,
which is for the most part self-corrected through the asymmetric drift term in Equation \ref{eq:vc}.
Instead, we attribute this difference between the simulation and the mocks is due to the selection function and limitations of the tracer population.

When we apply our quality cuts, we see that the red points in Fig. \ref{fig:aurigaiadiff} tend to be a closer match to the green line (the simulation particles)
for $R\lesssim15~\si{kpc}$.
The cut in magnitude has the largest impact on the results for $v_{\rm c}$ due to its impact on the selection function of the tracer population.
The next most impactful cut is that of selecting only giants.
For higher radii the complicated selection function and limited amount of datapoints increases the mismatch with $v_{\rm c,true}$ and the simulation particles.
We also see that observational uncertainties introduce only a small effect after all quality cuts have already been applied (blue points in Fig. \ref{fig:aurigaiadiff}).
For all simulated galaxies except Au16 and Au24 we also see that $v_{\rm c}$ for the mock samples tends to start declining more steeply than $v_{\rm c,true}$
around $R\gtrsim18~\si{kpc}$ because of selection effects on the sample,
even though these disks are embedded in extended dark matter halos that follow relatively well NFW profiles \citep{Callingham2020}.

\subsection{Satellite Perturbations}
Many of the features that we see in the velocity moments and in \vc{} are qualitatively similar to those seen in the simulation L2 discussed in Sec.~\ref{sec:simulations:Sagittarius}. They can again be attributed to the perturbations of satellites by considering the recent accretion history of the  \texttt{Auriga} galaxies. Since these galaxies have had a more complex, cosmological history than the idealised L2 simulation, they can show more complex structure, {due to, for instance, the recoiling and precessing of the disk or the asymmetric torques on the disk system caused by the unique merger history}:

\begin{itemize}
\item  Au6 suffers a  large accretion event $8~\si{Gyr}$ ago at a mass ratio with the host of $1:12$.
This satellite only fully disrupted $0.5~\si{Gyr}$ ago,
after orbiting the host halo with apocenters at $\sim40~\si{kpc}$ and pericenters at $\sim20~\si{kpc}$.
This is likely to have dynamically heated the stellar disk,
as can be seen in the profiles for the second moments of the velocity in Fig.~\ref{fig:aurigaiahalos}.
It's disk is atypical and does not follow the expected behaviour:
an exponential decline in $v_R$ and $v_z$ and an approximately constant profile for $v_{\phi}$ after $R>12~\si{kpc}$.
A significant part of this behaviour is sufficiently corrected for when calculating $v_{\rm c}$ with Eq.~\ref{eq:vc},
as seen in Fig. \ref{fig:aurigaiadiff}. 

\item Au16 has had a quiescent merger history over the past $4~\si{Gyr}$,
    and has not suffered significant recent perturbations to the dynamical equilibrium of the disk system.
    As a result, its derived $v_{\rm c}$ is close to $v_{\rm c,true}$.

\item Au21 has the largest difference between $v_{\rm c}$ and $v_{\rm c,true}$.
The galaxy accreted a satellite $5.5~\si{Gyr}$ ago with a mass ratio to the host of $1:15$,
that survives to present day, with a recent pericenter at $15~\si{kpc}$ only $0.5~\si{Gyr}$ ago.
This has caused significant perturbations to the disk, as evident in its velocity profiles.

\item Au23 has had a satellite which fully disrupted in the host only $2.2~\si{Gyr}$ ago, but has had a quiet merger history since.
The existence of the wiggles in the second moments and $v_{\rm c}$ within $R<15~\si{kpc}$ is {likely due to disequilibrium structure.}

\item Au24 had three satellites that fully disrupted $\sim2~\si{Gyr}$ ago, but has had a quiet merger history since then.
Hence we expect some disequilibrium in the disk system, and due to the longer dynamical timescale of the outer disk,
we can also explain the larger mismatch with $v_{\rm c,true}$ in this way.

\item Au27 has suffered 7 satellites of various infall times and mass ratios, and various distances from the disk system.
It is less clear which of these has likely had the largest effect on the disk dynamics and the results in Fig. \ref{fig:aurigaiadiff},
but Au27 is clearly being impacted by its satellite dynamics.

\end{itemize}

These results reaffirm our conclusion from Sec.~\ref{sec:simulations:Sagittarius} that a satellite or system of satellites interacting with a disk system,
combined with observational limits and a complicated selection function can cause a mismatch in the inferred $v_{\rm c}$ and $v_{\rm c,true}$ of $5-15\%$,
as well as a declining trend for large radii in a Jeans derived rotation curve. 

\section{Discussion} \label{sec:discussion}

As can be concluded from the results presented in Sections \ref{sec:rotation curve} on the analysis of the \textit{Gaia} data and of the simulations in \ref{sec:simulations:Sagittarius} and \ref{sec:simulations:auriga},
several intricacies underlie the practice of calculating the circular velocity curve through the application of Jeans equations under the assumption of equilibrium and axisymmetry.
In this section, we re-emphasize the caveats and complications that arise and which need further scrutiny.

\subsection{Sample Selection and Distances} \label{sec:dis:sampleselection}
Comparing our results to previous studies, we see slight differences in the behaviour of the radial profiles of the second moments of the various velocity components.
These differences are due to a combination of the specific survey used for the data, the sample selection, distance estimation,
and the assumed solar parameters. 

One selection we do not apply is a cut in $\left[\alpha/\text{Fe}\right]<0.12$ to remove contamination from the thick (high-alpha) disk,
as performed in \citetalias{Ou2023} using the APOGEE data of their sample.
We used the catalogue of calibrated metallicities from \citet{Recio-Blanco2023} to study the effect of the cut in $\left[\alpha/\text{Fe}\right]<0.12$,
but restricting our sample this way limited our data to $R<14~\si{kpc}$, and our results did not change within this radius.

In our study of the MW, we see a change in the behaviour of the radial profiles of all second velocity moments around $R\approx 12.5 - 15 ~\si{kpc}$,
in agreement with previous work \citep{Ou2023,Zhou2023,Jiao2023,Wang2023}.
This distance corresponds to a heliocentric distance of $\sim4 - 7 ~\si{kpc}$,
beyond which distance estimates typically have a larger dispersion when compared to stars that are closer to the Sun
\citep{Hogg2019,Anders2022,Bailer-Jones2023}.
While as we discuss below this is unlikely to be the reason behind the plateau in $\langle v_R^2\rangle$, our analysis of the simulations in Fig. \ref{fig:aurigaiadiff} shows that a distance quality cut can in fact produce an artificial decline in $v_{\rm c}$ at high radii even when $v_{\rm c,true}$ does not show evidence thereof. We refer the reader to \citet{Luri2018} for a more general discussion of distance quality and sample selection effects.

\subsection{Systematics on the Rotation Curve} \label{sec:dis:systematic}
As described and shown in Fig. \ref{fig:systematicsBJ},
the individual specific changes in the spatial selection or parameters of the model only affect the results systematically up to $2\%$ within $R<12.5~\si{kpc}$ and
$5\%$ up to $R<20~\si{kpc}$.
In the worst case, combining several of these systematics together can lead up to a $6\%$ effect for the inner radii and $15\%$ for the outer radii.
In the simulations however, both N-body and cosmological, we see that $v_{\rm c}(R)$ derived through Jeans equations can differ from $v_{\rm c,true}(R)$ up to $5-15\%$ as well.
Hence the methodological approach of applying the time-independent,
axisymmetric Jeans equations to a system showing signs of dynamical disequilibrium is certainly a source of mismatch.

For example, we find evidence of a departure from an exponential profile for $\langle v_R^2\rangle$ in the MW, in the sense of a flattening off for $R > 12.5$~kpc. This is also present in the data of \citet{Zhou2023} and  \citetalias{Ou2023}, but was instead interpreted as implying a larger scale-length, which is however not expected in steady-state configurations for systems with a constant scale-height (and recall that the MW's disk is known to be flared).
In Sec. \ref{sec:simulations:auriga}, we see that the plateau is more apparent for simulation particles than for the mock,
although we do observe a departure from the exponential profile around $R\approx12~\si{kpc}$ in most simulated galaxies.
From the simulations, we see that any bulk motion, satellite interaction,
(transient) spiral structure or other dynamical structure can impact this feature.
We conclude that the plateau is most likely due to a history of satellite interactions that caused non-equilibrium and asymmetric dynamical signatures.

Another systematic effect relates to the representation of the density of the tracer population. We have assumed an exponential density profile, but it has been suggested that the thin disk could be truncated around a radius of $R=15~\si{kpc}$ \citep[see e.g.][but see also \citealt{Lopez-Corredoira2018}]{Minitti2011}. Although we do not have evidence of a sharp truncation, we do see a sharp decrease in the numbers of stars at large radii (see the histogram in Fig. \ref{fig:radBJ}).
Additionally, the flare and the warp in the MW disk become important around $R\gtrsim12.5~\si{kpc}$,
and we saw the effect of the latter in Fig. \ref{fig:crosstermBJ}. {An important follow-up step is to include the effect of the warp on the modelling, like in \citet{Cabrera-Gadea2024} or \citet{Jonsson2024}.}
Furthermore, the assumed axis of symmetry of $z=0$ is not necessarily in the right position \citep{Chrobakova2022,Bovy2016}, and we see evidence of this also in the velocities' second moments.
Therefore, an important improvement could be made in the treatment of the spatial density of our tracer population, which enters $v_{\rm c}(R)$ through how we calculate the asymmetric drift correction term. We come back to this point below.

\subsection{A Keplerian decline of the Galactic rotation curve?}\label{sec:dis:keplerian}

A surprising result of recent work on the Galactic circular velocity curve is the claim of a Keplerian decline by \citet{Jiao2023}. 
Although not of similar amplitude, we do find a declining trend starting from $R\gtrsim18~\si{kpc}$ for the $v_{\rm c}(R)$ inferred using the Jeans equation for all halos except Au16,
even though these disks are embedded in a dark matter halo. Trying to fit a dark matter halo to such profiles
would require a more pronounced outer slope or cutoff,
and would likely result in a low virial mass.

\begin{figure}
    \includegraphics[width=0.9\linewidth]{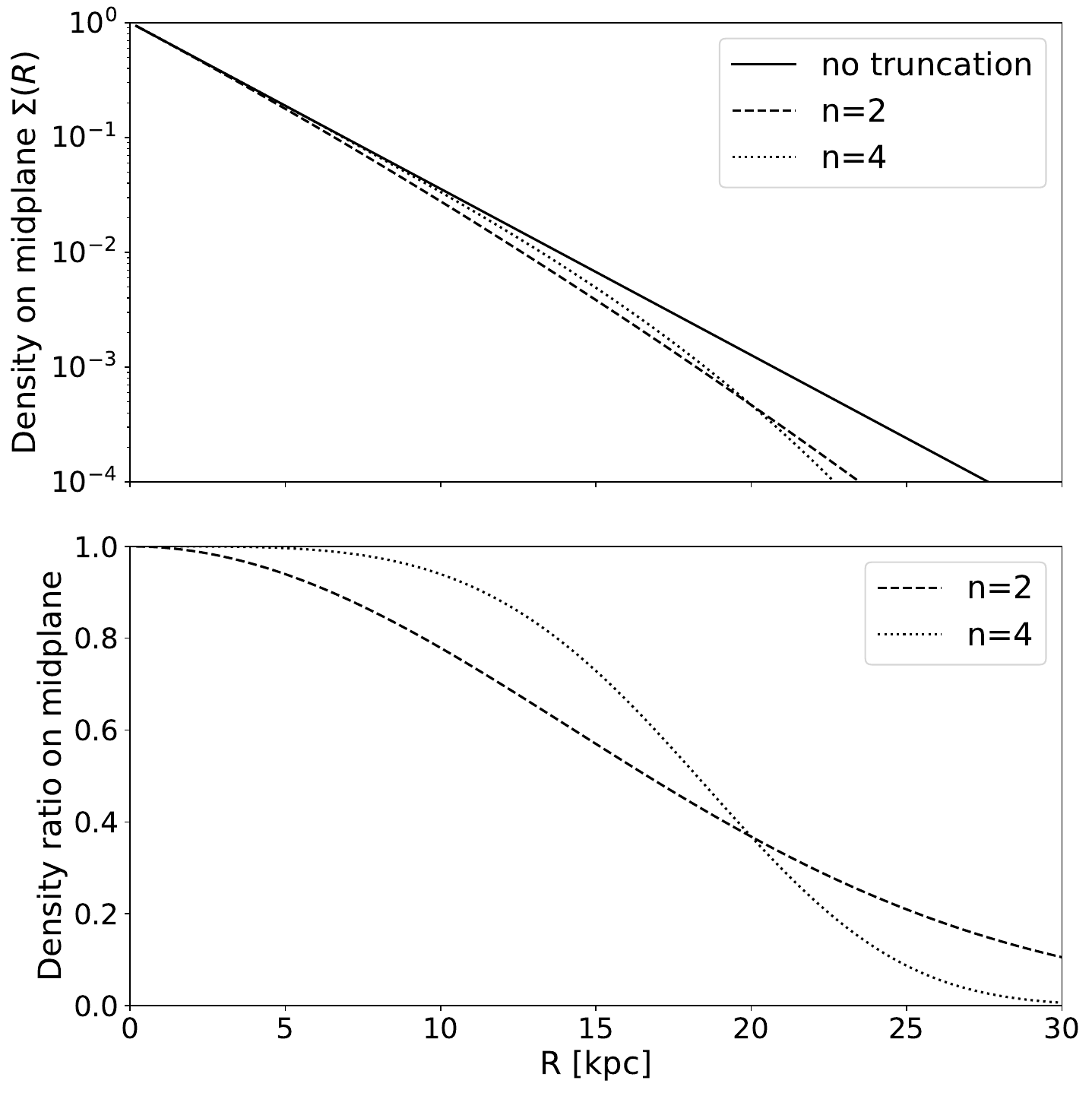}
    \includegraphics[width=0.8\linewidth]{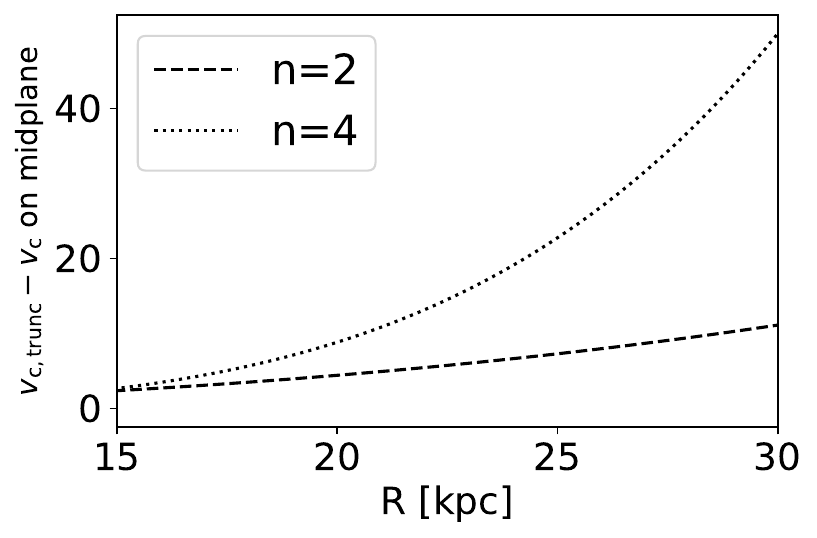}
    \caption{{\it Top}: Comparison of the density profiles from an exponential disk to two slowly truncated disk models. {\it Bottom:} Difference in the inferred circular velocity for the above models compared to an exponential disk.}
    \label{fig:delta_vc}
\end{figure}
Yet, the biggest source of uncertainty and possibly bias at larger distances lies in the density profile of the tracer population, particularly if this is affected by some kind of truncation. If we neglect the cross-term in Eq.~\ref{eq:vc}, the difference in the circular velocity that could arise when assuming a density profile $\nu_n$ compared to that obtained with the originally assumed $\nu$ is
\begin{equation}
v_{\rm c}^2 - v_{\rm c,n}^2 \equiv \Delta v_c^2 \sim - \langle v_R^2 \rangle \frac{\partial \ln \nu/\nu_n}{\partial \ln R}.
\end{equation}
In the top panels of Figure~\ref{fig:delta_vc} we compare the density profiles of the exponential disk $\nu$ from Eq.~\ref{eq:numdens} to two models of smoothly truncated disks $\nu_n = \nu \exp(-R^n/R_{\rm max}^n)$ for $n=2, 4$, with $R_{\rm max} = 20$~kpc. Since we can rewrite $\Delta v_{\rm c}^2=2v_{\rm c}\Delta v_{\rm c}-(\Delta v_{\rm c})^2$, then 
\begin{equation}
    \Delta v_{\rm c} \simeq 
        -\dfrac{n\langle v_R^2\rangle}{2v_{\rm c}} \dfrac{R^n}{R_{\rm max}^n},
\end{equation}
where we have assumed that $\Delta v_{\rm c}/v_{\rm c}$ is small.
The bottom panel of Fig.~\ref{fig:delta_vc} shows $\Delta v _{\rm c}$, assuming that $\sqrt{\langle v_R^2 \rangle} = 30$~km/s at $R>15~\si{kpc}$ and a linear decline for $v_{\rm c}(R) = v_{\rm c,\odot} - \beta(R-R_{\odot})$ as in \cite{Jiao2023}, 
where $v_{\rm c,\odot}$ is the circular velocity at $R_{\odot}$, the solar position, and $\beta=-2.18~\si{km.s^{-1}.kpc^{-1}}$.

Fig.~\ref{fig:delta_vc} thus reveals that the circular velocity inferred from the Jeans equation assuming a pure exponential profile could be underestimated by up to $\sim 10~\si{km.s^{-1}}$ at $30~\si{kpc}$ for $n=2$, and $\sim 45~\si{km.s^{-1}}$ for $n=4$, which is large enough to account for the steep decline seen for the MW \citep[see][]{Jiao2023,Wang2023}. We perform a slightly more in-depth study of this effect in Appendix \ref{app:trunc} and find again that modelling a truncated disk with an exponential density profile leads to an underestimated $v_{\rm c}$ in the outskirts. Therefore, this effect could potentially explain the declining rotation curve presented in other works and would leave room for a flat rotation curve for the Milky Way (as the decline is fully compensated by not accounting properly for the density profile of the population). It is thus of great importance to determine reliably the density profile of the outer Galactic disk.

\section{Summary and Conclusions} \label{sec:conclusion}

The main goal of this study was to investigate the validity of applying the radial Jeans equation under the assumption of steady-state and
axisymmetry to the MW to derive its circular velocity curve,
in particular given the recently reported signatures of disequilibrium in our Galaxy
\citep{Widrow2012,Williams2013,GaiaCollaboration2021anti,Antoja2018,VasilievTango2021,GaiaCollaboration2023Drimmel}.
We did this by studying the behaviour of the different terms in the axisymmetric Jeans equations in the MW stellar disk up to a galactocentric radius
 $R\approx21~\si{kpc}$ by using data from \textit{Gaia} DR3 \citep{GaiaCollaboration2023DR3}. 
We measured the second moments of the velocity ($\langle v_R^2\rangle$, $\langle v_{\phi}^2\rangle$, $\langle v_z^2\rangle$, $\langle v_Rv_z\rangle$)
throughout the MW disk.
Using these terms, we proceeded to derive the circular velocity curve $v_c$ for the MW, obtaining similar results as recent work.
Using two numerical simulations,
the N-body simulation L2 of a MW galaxy interacting with a Sagittarius dwarf \citep{Laporte2018} and the \texttt{Auriga} cosmological simulations \citep{Grand2017}, 
we verified our methodology.

Our analyses have led to the following findings.
\begin{itemize}
    \item The kinematics of disk stars in the MW show  vertical asymmetries and also with azimuth, providing further evidence that the MW is not in dynamical equilibrium (Figs. \ref{fig:azimBJ}, \ref{fig:radBJ}, \ref{fig:zpmBJ} and \ref{fig:zsplitBJ} and Sec.~\ref{sec:methods}).
     However, we find that applying Jeans equations to the MW using the methodology of this paper to calculate the circular velocity curve $v_{\rm c}$ is still robust
     up to $\sim5\%$ systematics for $R<14~\si{kpc}$ when considering only the data, i.e. intrinsic to the methodology.
     (Figs. \ref{fig:compvc} and \ref{fig:systematicsBJ} and Section \ref{sec:results:systematics}).
    Additionally, it is not significantly impacted by including the crossterm in Equation \ref{eq:vc} (Section \ref{sec:results:crossterm}).
    
    \item The radial profiles for $\langle v_R^2\rangle$ and $\langle v_z^2\rangle $ do not follow the expected exponential declining behaviour for a disk in equilibrium (i.e. with $R_{\rm exp,\langle v^2
_R \rangle} = 2 R_{\rm exp}$).
     This behaviour is also seen in simulated MW-like systems that have had interactions with satellites which enacted dynamical perturbations like spiral arms, flares or
      bulk motions across the stellar disk (Figs. \ref{fig:compBJSH}, \ref{fig:Laporteradialprofs} and \ref{fig:aurigaiadiff}
      and Sections~\ref{sec:rotation curve},~\ref{sec:simulations:Sagittarius},~\ref{sec:simulations:auriga}).
    
    \item 
    Such interactions can be the most important cause of
    the observed non-equilibrium signatures in the circular velocity curve derived through the radial Jeans equation
    (Figs. \ref{fig:aurigaiadiff}, \ref{fig:Laporteradialprofs}, \ref{fig:Laportetimeplots} and \ref{fig:aurigaiahalos} and Section \ref{sec:simulations:Sagittarius} and \ref{sec:simulations:auriga}).
     The mismatch with $v_{\rm c,true}$ is further impacted when considering observational limitations or a complicated selection function in the tracer population.
     From the simulations, we infer that these effects combined can cause a systematic of up to $15\%$ (Figs. \ref{fig:Laportetimeplots} and \ref{fig:aurigaiadiff}). 
     For the galaxy in the \texttt{Auriga} suite that resembles the MW the most (Au6), we find a systematic smaller than 10\% (which would translate into an error of $20\%$ in the enclosed mass). 
     \item Furthermore, ff the density of the tracer population were truncated at large radii but not be taken into account in the Jeans modelling, this could result in  a significant mismatch (e.g. of order 45 km~s$^{-1}$), leading to the impression of a fast declining rotation curve (\ref{sec:dis:keplerian}).
\end{itemize}

\noindent 

We have thus uncovered similar problems in modelling the mass content of our Galaxy using the Jeans equations to the works of \citet{Kafle2018}, \citet{ElBadry2017}, \citet{Wang2018} and \citet{Kretschmer2021}
for the spherical case, but now for the axisymmetric case \citep[see also][]{Chrobakova2020}. Nonetheless to first order, the radial Jeans equation proves to be a useful tool for deriving the circular velocity curve for the MW. However, with the present-day quality of the data, we now see the impact of higher-order effects and systematics resulting from simplifications and broken assumptions. We thus urge for caution when interpreting $v_{\rm c}$ values calculated through Jeans equations with the current methodology outside of $R>15~\si{kpc}$, as this region is particularly affected by asymmetries, the warp and flare of the MW disk and kinematic disturbances (Section \ref{sec:discussion}).

The systematics associated with the data sample could be reduced by more, and better, measurements of the kinematics of disk stars at large distances
The most significant improvements will likely come from more flexible modelling methodologies.
It will be important to investigate possibilities of abandoning the assumptions of time-independence and axisymmetry
to derive the Galactic rotation curve with higher precision.
%---------------------------------------------
%    Acknowledgements, facilities, software
%---------------------------------------------

\begin{acknowledgements}
We thank Robert Grand who kindly provided clarifications regarding the \texttt{Auriga} simulations used in this work. OK also thanks Christine Eilers and Xiaowei Ou for sharing their data and insights with us, as well as Friedrich Anders and Eugene Vasiliev for helpful discussions. We are grateful to Filippo Fraternali for scientific interactions and literature suggestions. We~acknowledge~financial~support from a Spinoza prize to AH. The work reported on in this publication has been partially supported by COST Action CA18104: MW-Gaia.
TA acknowledges the grant RYC2018-025968-I funded by MCIN/AEI/10.13039/501100011033 and by ``ESF Investing in your future’’, the grants PID2021-125451NA-I00 and CNS2022-135232 funded by MICIU/AEI/10.13039/501100011033 and by ``ERDF A way of making Europe’’, by the ``European Union'' and by the ``European Union Next Generation EU/PRTR'' and the Institute of Cosmos Sciences University of Barcelona (ICCUB, Unidad de Excelencia ’Mar\'{\i}a de Maeztu’) through grant CEX2019-000918-M.
%Gaia acknowledgement 
This work has made use of data from the European Space Agency (ESA) mission
{\it Gaia} (\url{https://www.cosmos.esa.int/gaia}), processed by the {\it Gaia}
Data Processing and Analysis Consortium (DPAC,
\url{https://www.cosmos.esa.int/web/gaia/dpac/consortium}). Funding for the DPAC
has been provided by national institutions, in particular the institutions
participating in the {\it Gaia} Multilateral Agreement.

% packages
Throughout this work, we have made use of the following packages: \texttt{astropy} \citep{Astropy},
          \texttt{vaex} \citep{vaex2018},
          \texttt{SciPy} \citep{2020SciPy-NMeth},
          \texttt{matplotlib} \citep{matplotlib},
          \texttt{NumPy} \citep{Numpy},
          \texttt{GADGET-4} \citep{GADGET-4},
          \texttt{AGAMA} \citep{AGAMA} and Jupyter Notebooks \citep{JupyterNotebook}.
\end{acknowledgements}

%---------------------------------------------
%                 Bibliography 
%---------------------------------------------
\bibliography{bibliography}
\bibliographystyle{aa} 

%---------------------------------------------
%                 Appendix
%---------------------------------------------

\begin{appendix}

\section{Figures for the StarHorse sample} \label{app:shfigs}
For readability of the main manuscript we present the distributions for the SH sample here, since they agree qualitatively with the results on the BJ sample.
\begin{figure}[ht]
    \centering
    \includegraphics[width=\linewidth]{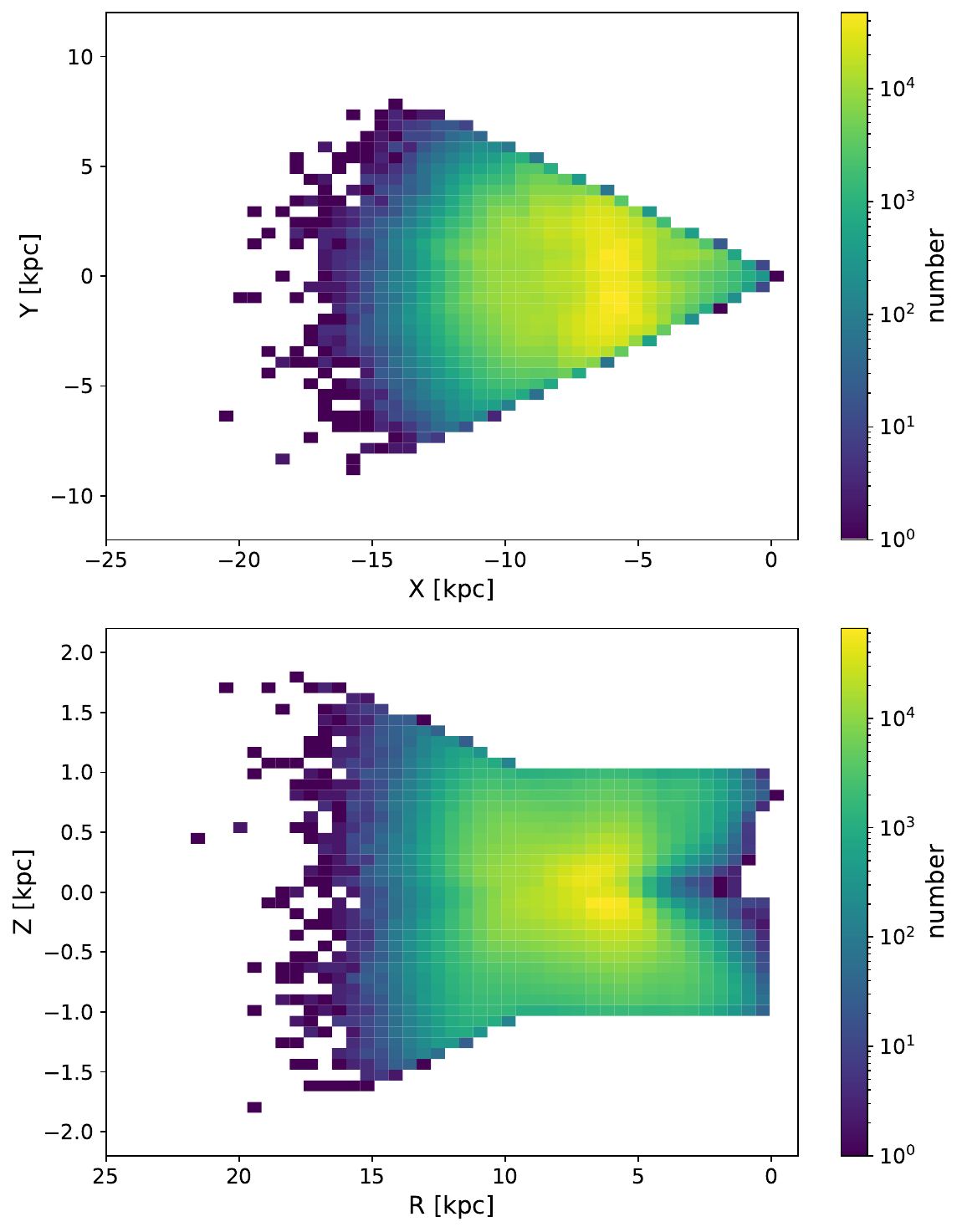}
    \caption{Same as Figure \ref{fig:spatialsample} but for the SH sample.}
    \label{fig:spatialsampleSH}
\end{figure}
\begin{figure}[ht]
    \centering
    \includegraphics[width=\linewidth]{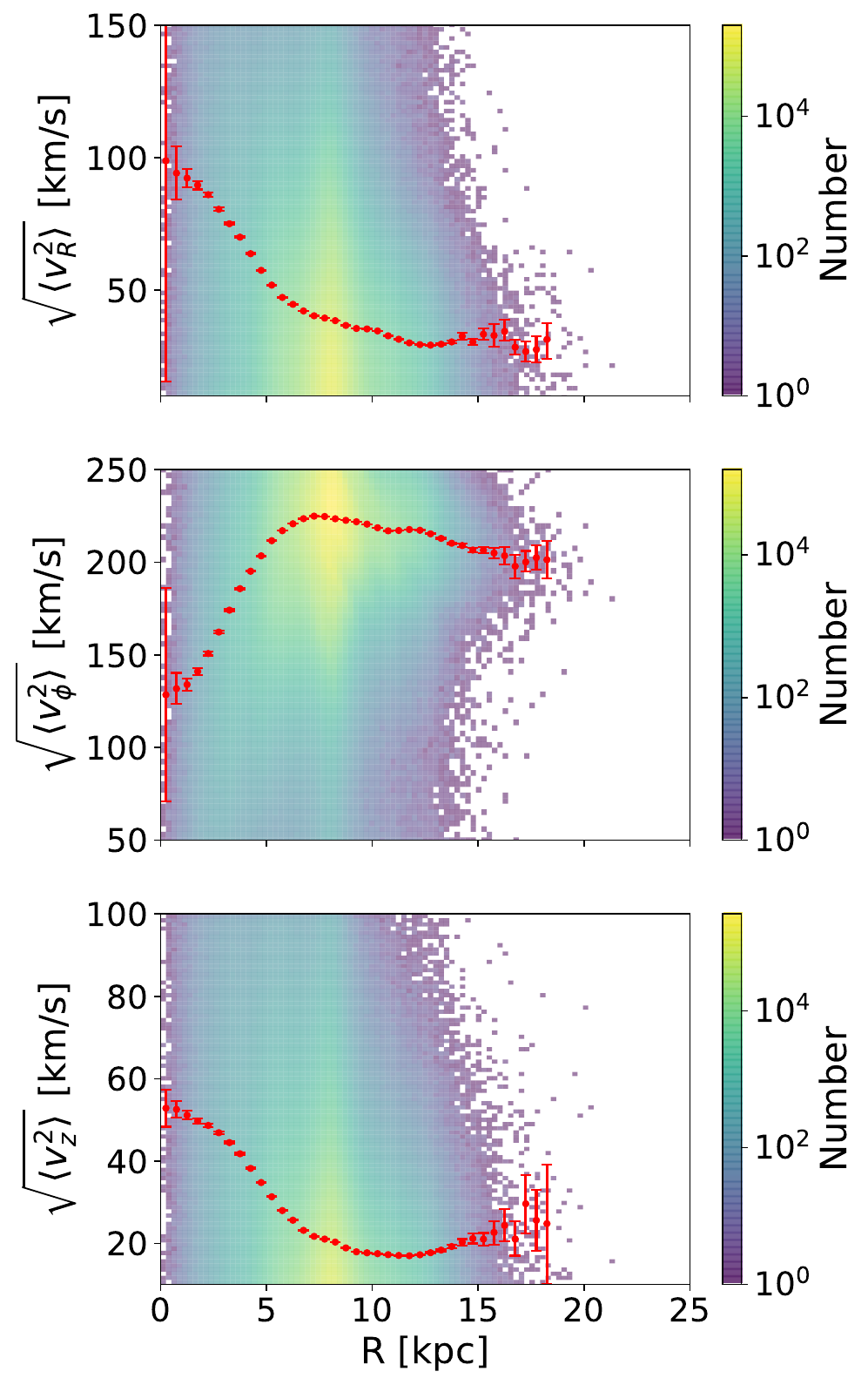}
    \caption{Same as Figure \ref{fig:radBJ} but for the SH sample.}
    \label{fig:radSH}
\end{figure}
\begin{figure*}[ht]
    \centering
    \includegraphics[width=\textwidth]{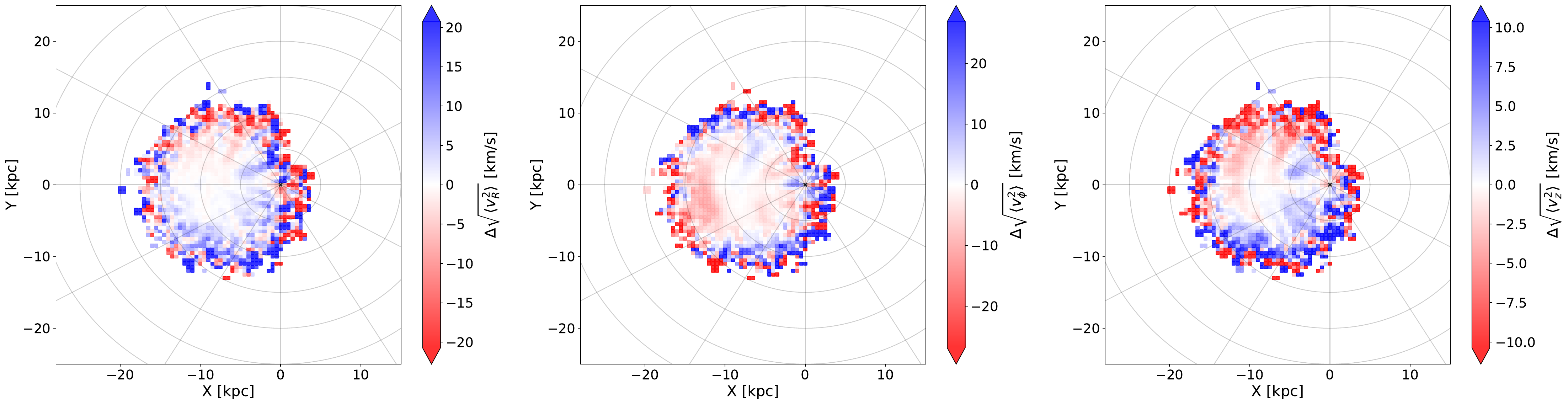}
    \caption{Same as Figure \ref{fig:zpmBJ} but for the SH sample.}
    \label{fig:zpmSH}
\end{figure*}
\begin{figure}[ht]
    \centering
    \includegraphics[width=\linewidth]{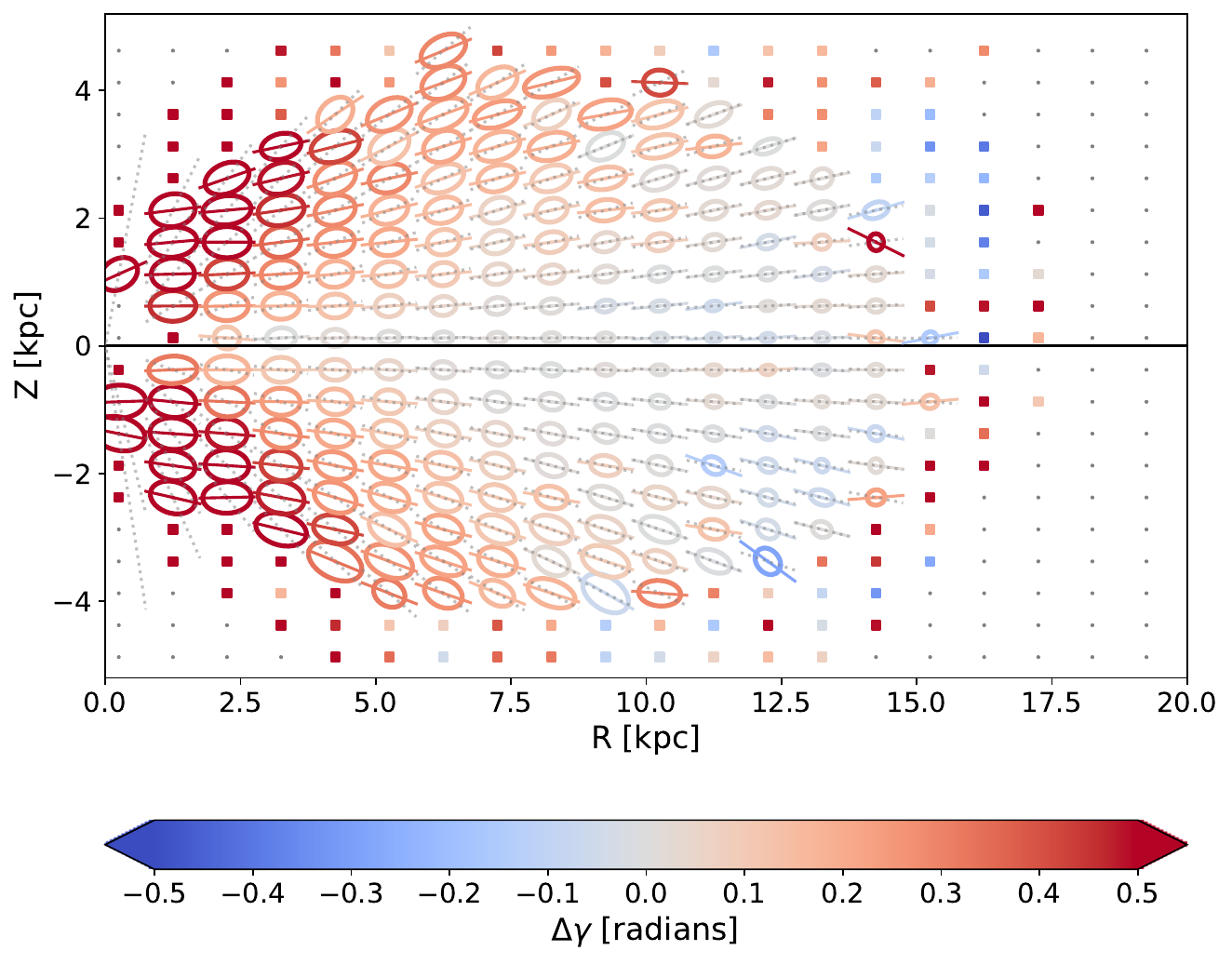}
    \caption{Same as Figure \ref{fig:tiltBJ} but for the SH sample.}
    \label{fig:tiltSH}
\end{figure}

\section{Azimuthal split for the Bailer-Jones sample}\label{app:azfigs}

\begin{figure}[ht]
    \centering
    \includegraphics[width=\linewidth]{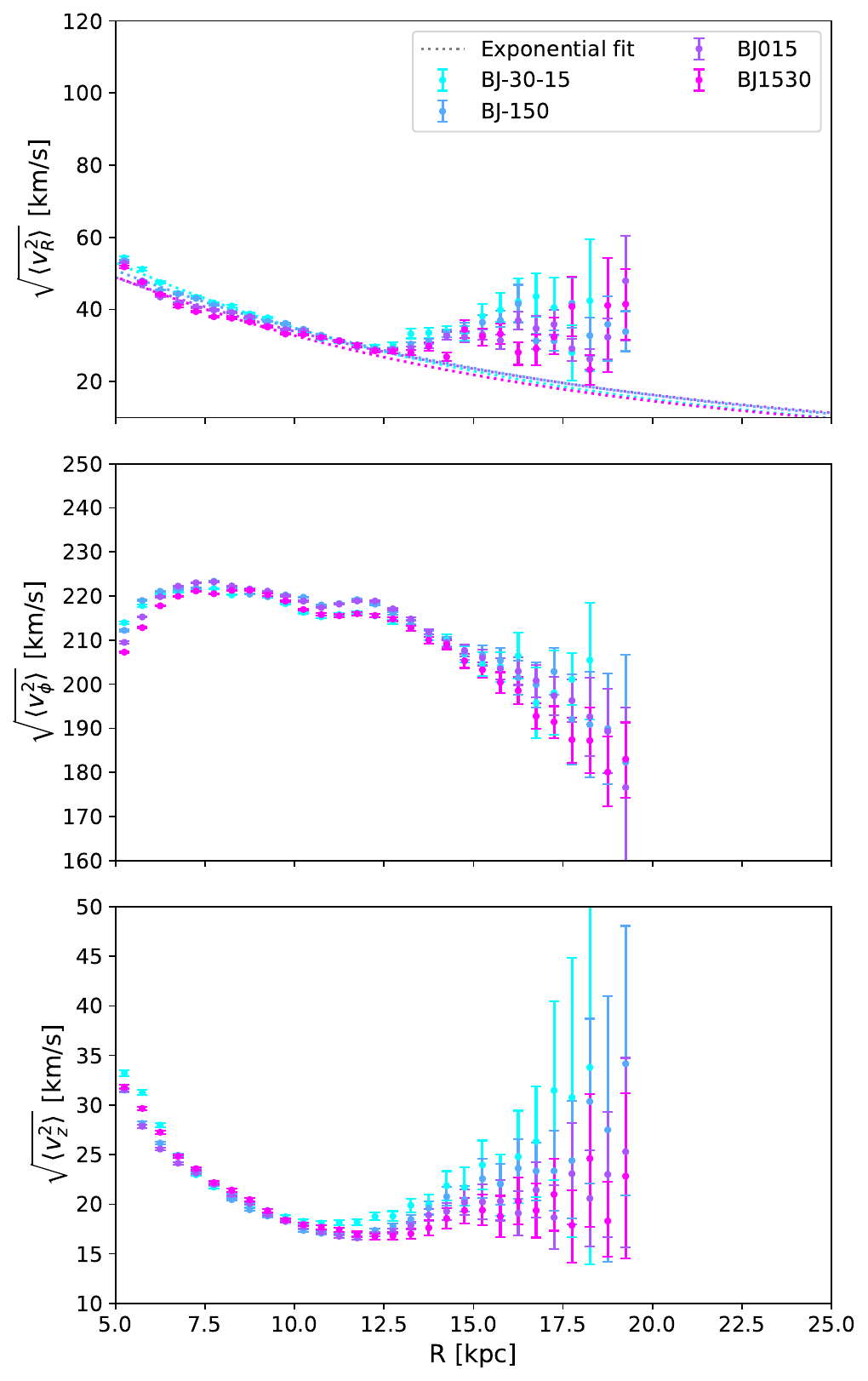}
    \caption{Comparison of the three radial profiles of the second moments of the velocity distributions in the Milky Way stellar disk for the BJ sample split into four smaller azimuthal slices between $(-30^{\circ},-15^{\circ},0^{\circ},15^{\circ},30^{\circ})$.}
    \label{fig:azimBJ}
\end{figure}

Figure \ref{fig:azimBJ} shows the radial profiles for the BJ sample split up into azimuthal slices of $15^{\circ}$.
We can see how there appears to be a trend in $\langle v_R^2\rangle$ and $\langle v_z^2\rangle$ in that they decrease overall with increasing $\phi$, seen by the lightblue points generally lying above the darker blue, purple and pink points in the leftmost and rightmost panels of Fig. \ref{fig:azimBJ}. Furthermore, the azimuthal second moment seems to be slightly lower for the wedges further away from the anticenter, and higher for those at lower azimuths, which is likely to be a selection effect due to poor sampling. We included two lines in Fig. \ref{fig:systematicsBJ} showing the effect of taking $30^{\circ}$ wedges to avoid selection effects, and can conclude that there is asymmetry in azimuth in the Milky Way disk.

\section{Aurigaia mock particles}\label{app:mocks}
As described in \cite{Grand2018aurigaia} and \cite{Lowing2015}, to generate the \textit{Gaia} DR2 mock catalogues for the six \texttt{Aurgaia} halos, a population of stars is generated for each particle in the final simulation snapshot. As seen in Fig. \ref{fig:aurigaiadiff}, the mock sample for a given Auriga halo will typically have a larger $\Delta v_{\rm c}$ than the sample with simulation particles. In this appendix we investigate what is the driving force behind this larger $\Delta v_{\rm c}$.
We present results for studying this effect only in halo Au6, but similar findings hold for the other halos.

For each mock star, the particle ID of the simulation particle that it was spawned from is stored in the mock catalogue. We call this simulation particle the `parent particle'. Thus we can find the sample of parent particles that have generated the mock particles in our selected sample. Typically, except for the region of $\approx1.5~\si{kpc}$ around the solar position, the parent samples contain fewer particles than the simulation particle sample discussed in Section \ref{sec:simulations:auriga}. Near the Sun, $v_{\rm c}$ for parent particles is equal to the $v_{\rm c}$ found for the simulation particles. Also, in any region further away from the Sun, some of the simulation particles in the parent sample are not within our spatial selection, but end up in the sample because they have generated a mock particle that did end up in our selection. These parent particles inflate the tails of the velocity distribution of particles in the relevant radial bins, hence inflating the dispersion and, thus, the second moments of the velocity distributions as well. That is why we opt to minimize this numerical effect by keeping only mock stars that were generated from parent stars within our spatial selection.

\subsection{Effect of generating a mock sample}\label{app:mocks:gen}
\begin{figure}[ht]
    \centering
    \includegraphics[width=\linewidth]{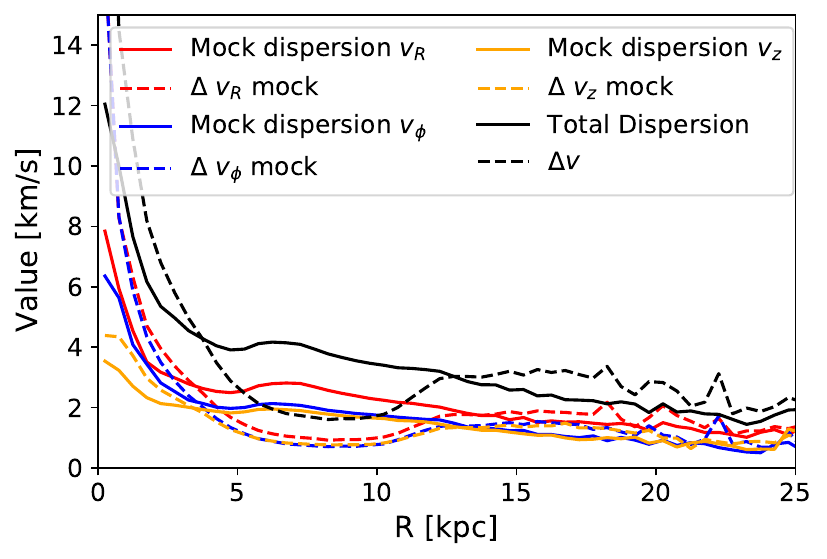}
    \caption{Radial profiles of the average dispersion and displacement of the mock particles with respect to their parent particles. Black lines show the total over all three velocity components. Solid (dashed) lines show the dispersion (displacement). Red, blue and green lines show the values for $v_R$, $v_{\phi}$ and $v_z$ respectively.}
    \label{fig:parentdispersion}
\end{figure}

A final question can be asked about the generating of mock particles from the simulation particles and if it causes some artificial dispersion to our mock sample. To investigate this artificial dispersion, we start by finding all parent particles of stars in our mock sample. Then, for each parent particle, we look at all mock stars spawned from that particle that are present in our mock. We calculate the dispersion of this group of stars, and their mean velocity. This mean velocity often differs from the velocity of the parent particle, creating a non-zero $\Delta v$. Figure \ref{fig:parentdispersion} shows the radially binned and averaged profiles of these dispersions and displacements.

We see here that we can expect an artificial mean displacement of particles in velocity of about $3~\si{km/s}$, and a dispersion of about $4~\si{km.s^{-1}}$. This is contrary to the quite good match within $2\%$ between the mock and simulation results (the blue points and green dashed line, respectively) in Fig. \ref{fig:aurigaiadiff}. When calculating $v_{\rm c}$, the artificial dispersion and displacement in $v_{\phi}$ get mostly self-corrected through the asymmetric drift term (the second term in Equation \ref{eq:vc}), which is why this does not influence our result for $v_{\rm c}$ more than $1-2\%$. The additional dispersion from generating the mock particles does explain part of the mismatch between the radial profiles for the same samples in Fig. \ref{fig:aurigaiahalos}. Still, we find that the sample selections and selection function are more important in causing the differences seen in Fig. \ref{fig:aurigaiadiff}.

\section{Action Distribution Models of Truncated Discs}\label{app:trunc}

\begin{figure}[t]
    \centering
    \includegraphics[width=\linewidth]{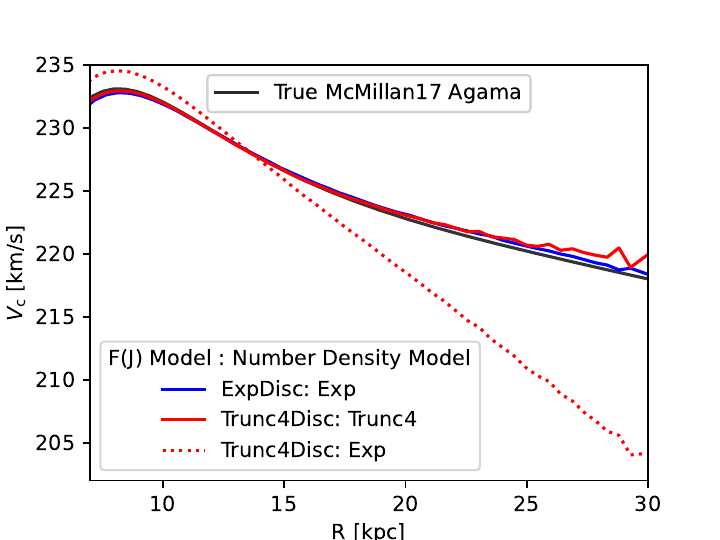}
    \caption{The results of applying Jeans Modelling to find the velocity curves from samples drawn from action distributions.
            We create samples of tracer particles drawn from three different models,
            a model representing an approximately exponential disk,
            and two models where the disk has been truncated, with the Trunc4 model sharply truncated.
            The solid lines correspond to velocity curves found using number densities that correctly match the action distribution model.
            The dotted lines show the results of incorrectly assuming that the truncated samples follow an exponential number density.
    }
    \label{fig:FJ}
\end{figure}

To test the effects of a truncated disk, we consider a set of toy models,
generated by action distributions in a fixed potential using \textsc{agama}  \citep{AGAMA}.
We model the thin disk using the QuasiIsothermal DF, which gives a number density profile and radial velocity dispersion profile that are very close to exponential.
We use the following parameters: $R_\mathrm{disk}=3\kpc$, $H_{\mathrm{disk}}=0.2\kpc$,  $\sigma_{r,0}=75\kms$,  and $R_{\sigma, r}=25\kpc$.
Note that $R_{\sigma,r}$ corresponds to the exponential scale length of $\sigma_{v_{R}}$,
and is equivalent to twice that of  $R_{\rm exp, \langle v_R^2\rangle}$.

To model truncated disks, we modify the QuasiIsothermal DF as:
\begin{equation}
    \exp\left(-\frac{R_\mathrm{c}}{R_{\mathrm{disk}}}\right) \rightarrow
\exp\left(-\frac{R_\mathrm{c}}{R_{\mathrm{disk}}} - {\left(\frac{R_\mathrm{c}}{R_{\mathrm{max}}}\right)}^{n} \right) 
\end{equation}
where $R_{\mathrm{max}}$ is the truncation radius, and $n$ gives the steepness of the trunctation.
We use $R_{\mathrm{max}}=20$ and $n=4$.

From these distributions, we draw positions and velocities of $10$ million tracer particles,
that follow physical selections of $6\kpc<R<25\kpc$ and $|z|<0.1\kpc$.
These particles are purely a tracer population and has no effect on the underlying galactic potential, which is fixed to the potential of \citet{McMillan17}.
Upon these particles, we perform the same modelling as we apply to the MW.
We fit exponential profiles to the radial velocity, finding that they approximately follow the parameters of the distribution function.
The crossterm is approximated using a dataset with a relaxed physical selection of $|z|<1\kpc$.

For the number density profile, we model the truncated discs with both the true corresponding truncated number density profile,
and a best-fit exponential profile.
The results can be seen in Fig.~\ref{fig:FJ}.
When the true number density of the tracer population is known, the velocity curve is well recovered (solid curves), and closely matches the true value of the potential.
The small differences between the true velocity curve and the correct models can be attributed to the small difference between the true number density generated by the action distribution models and the analytic exponential and truncated exponential profiles.

However, if the tracer population is truly distributed as a truncated profile and is instead modelled as exponential, strong systematic effects appear.
The recovered velocity curve has a noticeable decline, with a slope of $\beta\approx-1.5 \si{km.s^{-1}.kpc^{-1}}$.
A sharper truncation, with larger $n$, or a smaller $R_\mathrm{max}$, constitute a stronger decline.

\end{appendix}

%---------------------------------------------
\end{document}